\documentclass[%
 jor,
 prl,
 amsmath,amssymb,
twocolumn,
]{revtex4-2}

\usepackage{graphicx}
\usepackage{dcolumn}
\usepackage{bm}
\usepackage{bbold}
\usepackage{float}

\renewcommand{\vec}[1]{\bm{\mathbf{#1}}}

\begin{document}

\title{Electro-nuclear transition into a spatially modulated magnetic state in YbRh\textsubscript{2}Si\textsubscript{2}}

\author{J.\ Knapp}
\affiliation{
Department of Physics, Royal Holloway University of London, TW20 0EX, Egham, UK.
}
 \email{jan.knapp@rhul.ac.uk, l.v.levitin@rhul.ac.uk, j.saunders@rhul.ac.uk}
\author{L.\ V.\ Levitin}
\affiliation{ 
Department of Physics, Royal Holloway University of London, TW20 0EX, Egham, UK.
}
 \email{l.v.levitin@rhul.ac.uk}
\author{J.\ Ny\'eki}
\affiliation{ 
Department of Physics, Royal Holloway University of London, TW20 0EX, Egham, UK.
}
\author{A.\ F.\ Ho}
\affiliation{ 
Department of Physics, Royal Holloway University of London, TW20 0EX, Egham, UK.
}
\author{B.\ Cowan}
\affiliation{
Department of Physics, Royal Holloway University of London, TW20 0EX, Egham, UK.
}
\author{J.\ Saunders}
\affiliation{
Department of Physics, Royal Holloway University of London, TW20 0EX, Egham, UK.
}
 \email{j.saunders@rhul.ac.uk.}
\author{M.\ Brando}%
\affiliation{ 
Max Planck Institute for Chemical Physics of Solids, N\"othnitzer Stra\ss{}e 40, 01187 Dresden, Germany.
}
\author{C.\ Geibel}%
\affiliation{ 
Max Planck Institute for Chemical Physics of Solids, N\"othnitzer Stra\ss{}e 40, 01187 Dresden, Germany.
}

\author{K.\ Kliemt}%
\affiliation{ 
Physikalisches Institut, Max-von-Laue-Stra\ss{}e 1, 60438 Frankfurt am Main, Germany.
}
\author{C.\ Krellner}%
\affiliation{ 
Physikalisches Institut, Max-von-Laue-Stra\ss{}e 1, 60438 Frankfurt am Main, Germany.
}

\date{9 January 2023}

\begin{abstract}
The nature of the antiferromagnetic order in the heavy fermion metal YbRh\textsubscript{2}Si\textsubscript{2}, its quantum criticality, and superconductivity, which appears at low mK temperatures, remain open questions. We report measurements of the heat capacity over the wide temperature range 180\,$\mu$K - 80\,mK, using current sensing noise thermometry.   In zero magnetic field we observe a remarkably sharp heat capacity anomaly at 1.5\,mK, which we identify as an electro-nuclear transition into a state with spatially modulated electronic magnetic order of maximum amplitude 0.1\,$\mu_B$. We also report results of measurements in magnetic fields in the range 0 to 70\,mT, applied perpendicular to the c-axis, which show eventual suppression of this order. These results demonstrate a coexistence of a large moment antiferromagnet with putative superconductivity.\\[-1.4em]\rule{\textwidth}{0pt}
\end{abstract}

\maketitle

The interplay of magnetism and superconductivity is a central question in the study of strongly correlated electronic systems. In heavy fermion (HF) metals a particular advantage is the ability to tune the system to a quantum critical point (QCP), by pressure or some other tuning parameter, at which superconductivity can emerge. In YbRh\textsubscript{2}Si\textsubscript{2} magnetic field provides a convenient tuning parameter, at ambient pressure, and without recourse to doping. However superconductivity in YbRh\textsubscript{2}Si\textsubscript{2} only appears at low mK temperatures, implying extremely low thermodynamic critical fields. The onset of strong magnetic screening and a heat capacity peak observed in the vicinity of 2\,mK~\cite{Schuberth2016} have been interpreted in terms of a simultaneous superconducting and electro-nuclear magnetic phase transition. The experiment we report in this Letter focuses on a detailed and precise investigation of this transition, on establishing the magnetic ground state, and its evolution with magnetic field.  

YbRh\textsubscript{2}Si\textsubscript{2} has tetragonal symmetry and a theoretically predicted highly anisotropic, three dimensional Fermi surface \cite{Friedemann2010b,Kummer2015,Zwicknagl2016,Li2019a,Guettler2021}. Antiferromagnetic (AFM) electronic order appears in zero applied field at $T_N=70$\,mK and features ultra-small ordered moments, $\mu_{e}\approx0.002\,\mu_B$ \cite{Ishida2003}, which develop out of partially Kondo-screened Yb local moments $1.4\mu_B$~\cite{Gegenwart2002}.
The nature of this order is not established, with an interesting possibility of the ordered moments aligned with the magnetically-hard c-axis~\cite{Hamann2019}.
Neutron scattering, above $T_N$, shows incommensurate AFM fluctuations which emerge from ferromagnetic (FM) fluctuations \cite{Stock2012}.
Static magnetic susceptibility \cite{Gegenwart2005}, NMR \cite{Ishida2002} and ESR \cite{Sichelschmidt2003,Abrahams2008,Woelfle2009} also provide evidence of FM fluctuations.

The observed suppression of $T_N$ by magnetic field at ambient pressure on high quality samples first led to the proposal of a QCP, induced by an in-plane field of $B_c = 60$\,mT, or ten times larger field along the c-axis \cite{Gegenwart2002}, reflecting the highly anisotropic electronic magnetism.  The nature of the putative QCP remains a matter of debate, including theories of local quantum criticality \cite{Si2001,Coleman2005,Si2010,Steglich2014}, see also \cite{Schubert2019,Gegenwart2007,Gegenwart2008,Paschen2020,Schuberth2022} and theories invoking strong coupling of fermions and spin fluctuations into critical quasiparticles \cite{Abrahams2012,Abrahams2014,Woelfle2015}. Negative chemical pressure, achieved by Ge doping, shifts the QCP to smaller fields \cite{Custers2003a,Gegenwart2005}, cobalt doping induces ferromagnetism \cite{Lausberg2013,Hamann2019}.

Most recently, the report of superconductivity in YbRh\textsubscript{2}Si\textsubscript{2} \cite{Schuberth2016,Nguyen2021} led to the proposal that an important role is played by the coupling of electronic and nuclear magnetism. The strong hyperfine interaction and presence of active Yb nuclei distinguishes YbRh\textsubscript{2}Si\textsubscript{2} from Ce-based HF systems, for which the nuclear moments are zero. Thus YbRh\textsubscript{2}Si\textsubscript{2} provides a model system to investigate the influence of nuclear spins in a Kondo lattice exhibiting quantum criticality \cite{Eisenlohr2021}. The ground state doublet of the Yb ion in the crystalline electric field (CEF), also distinguishes this system from systems with strong hyperfine interactions based on non Kramers ions such as Pr and Ho \cite{Libersky2021}. The work reported here presents a first step to precisely thermodynamically characterize the interplay of electronic and nuclear magnetism in YbRh\textsubscript{2}Si\textsubscript{2}.

\begin{figure}[t!]
    \centering
    \includegraphics{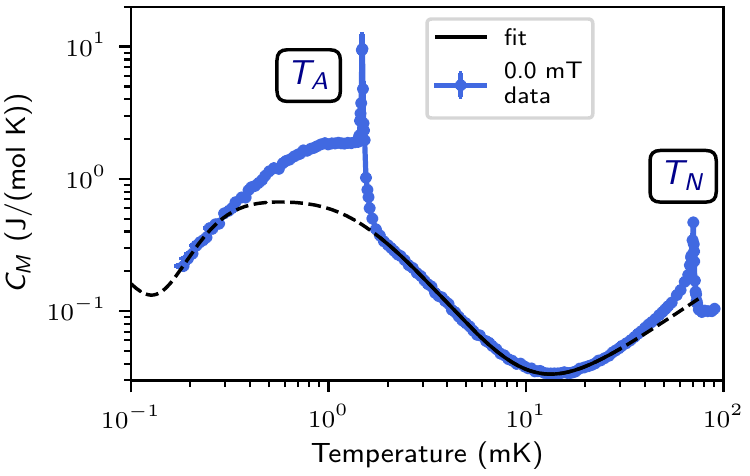}
    \caption{Molar heat capacity in zero field exhibits two sharp anomalies at $T_A$ and $T_N$ that we identify with magnetic transitions. The data between 1.85 and 30\,mK are fitted to the nuclear and heavy-electron heat capacity.
    The fit curve is plotted outside of the fitting interval as a dashed line.
    The small ordered electronic moments found above $T_A$~\cite{Ishida2003} would significantly affect the nuclear heat capacity only below 0.2\,mK.
    }
    \label{fig:Zero_field_fit}
	\vskip1em
    \centering
    \includegraphics{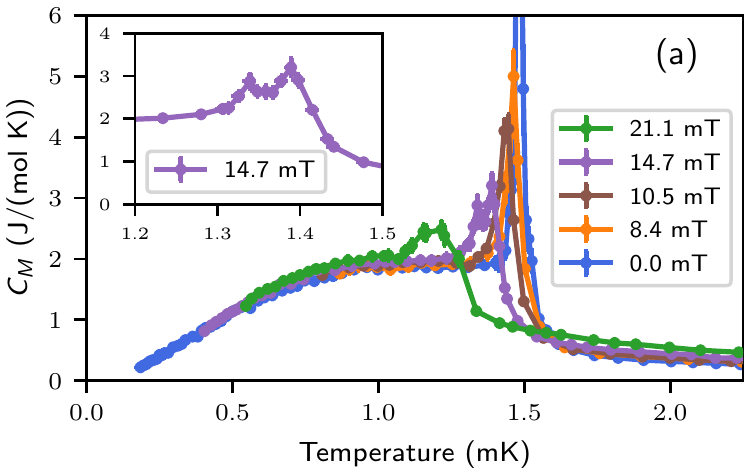}
    ~\includegraphics{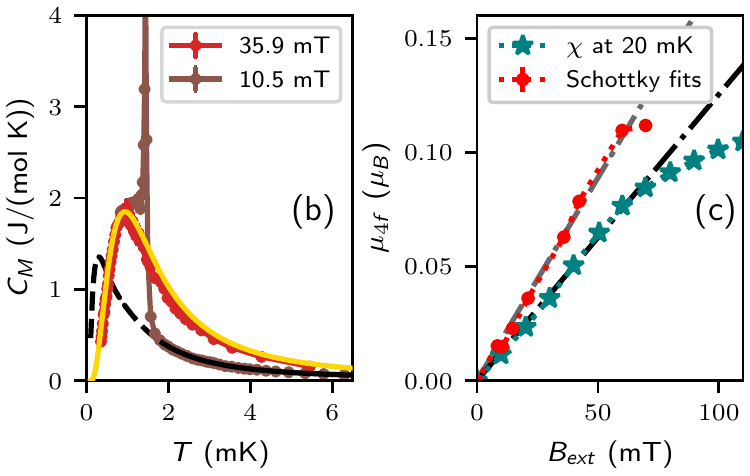}
    \caption{Molar heat capacity in field applied in the ab-plane.
    (a) Suppression of $T_A$ anomaly with field.
    (b) Examples of fitting the Schottky model. Below 35.9 mT, where $T_A$ is observed, only the data above this anomaly are fitted.
    (c) The static electronic moment of Yb determined from ac susceptibility $\chi$ \cite{Brando2013} and from the Schottky model.}
    \label{fig:field_data}
\end{figure}

Our experimental set-up exploits advances in current sensing noise thermometry~\cite{Casey2014}. This includes improvements in the speed of measurement achieved by a relatively-high sensor resistance (a 0.2\,$\Omega$ PtW wire),
coupled with the ability to limit the heat leak into the noise thermometer to below 1\,fW by appropriate shielding and filtering of the leads \cite{Levitin2022a}. 
The single crystal of YbRh\textsubscript{2}Si\textsubscript{2} from batch 63129 with $RRR = 50$~\cite{Krellner2012} is thermalised via an aluminium wire, operating as a superconducting heat switch. A superconducting solenoid both provides the sample field and operates the heat switch. The heat capacity is determined by the adiabatic heat pulse method below 10\,mT, the critical field of aluminium,
and by the relaxation method above it.

The molar heat capacity in zero applied field, Fig.~\ref{fig:Zero_field_fit}, shows the well-known N\'eel anomaly at $T_N = 70.5$\,mK, and another sharp anomaly at $T_A=1.5$\,mK.
The heat capacity measured around 1\,mK exceeds the heavy-electron term $\gamma_S T$ by 3 orders of magnitude. We demonstrate that this large heat capacity originates from Yb nuclear degrees of freedom, however the $T_A$ anomaly reflects a cooperative transition involving both nuclei and electrons.
On the other hand, above a few mK the nuclear heat capacity decreases as $T^{-2}$, while the electronic heat capacity increases linearly with temperature.
As a result, above 20\,mK the electronic part dominates (see SI, Fig. 7), and thus at $T_N$ the nuclear spin degrees of freedom play no role \cite{Knebel2006}, contributing less than 1\% to the heat capacity there.
While the electronic moments form a regular lattice, only a minority of Yb sites 
carry a nuclear moment. Thus the low temperature heat capacity arises from the nuclei of \textsuperscript{171}Yb ($I=1/2$) and \textsuperscript{173}Yb ($I=5/2$) isotopes with 0.1431 and 0.1613 natural abundances respectively, distributed randomly across Yb sites.
The nuclear spins are subject to an effective hyperfine magnetic field $\vec B_{hf} = -A_{hf} \vec{\mu}_{e}$,
produced by the ordered static part of Yb electronic moments $\vec{\mu}_{e}$.
Here $A_{hf} = 102\,\mathrm{T}/\mu_B$ is the hyperfine constant \cite{Kondo1961,Bonville1984,Bonville1991}. The ``fast relaxation regime'', realised in YbRh\textsubscript{2}Si\textsubscript{2} \cite{Nowik1968,Plessel2003,Knebel2006,Flouquet1975,Flouquet1978},
enables us to ignore the hyperfine field due to the fluctuating part of the electronic moments and treat $\vec{\mu}_{e}$ as a mean field.
Additionally the $^{173}$Yb nuclei experience quadrupolar splitting in the crystalline electric field gradient, that points along the c-axis.


We neglect interactions between nuclei and consider a single-spin Hamiltonian
\begin{equation}
    \begin{gathered}
    \widehat{H} = g\mu_N A_{hf} \vec{\widehat{I}}\cdot \vec{\mu}_{e} + \frac{e^2qQ}{4I(2I-1)}(3\widehat{I}_z^2 - I(I+1)),
    \end{gathered}
\label{eq:Hamilton}
\end{equation}
where $g$, $\mu_N$, $Q$ are the nuclear g-factor, magneton and quadrupole moment and $eq$ represents the electric field gradient.
In general the nuclear spin $\vec{\widehat{I}}$ is not aligned with the c axis and
Eq.~\eqref{eq:Hamilton} is diagonalised numerically, to calculate the partition function $Z$ of the nuclear system, and hence thermodynamic quantities.
These are summed over a random distribution of $^{171}$Yb and $^{173}$Yb nuclei according to their natural abundance.

In the simplest case, the \emph{Schottky model}, we assume uniform $\mu_{e}$ on all Yb sites.
Fitting the data above $T_A$ unambiguously proves that the size of the static electronic moment in zero magnetic field is small, in agreement with measurements of muon spin resonance \cite{Ishida2003}. We put an upper bound $\mu_{e}\ll0.01\,\mu_{B}$ (in any direction) and directly determine the parameters of the quadrupole splitting.
We find a positive electric field gradient $eq=(2.06\,\pm\,0.01)\cdot10^{21}$\,Vm\textsuperscript{-2}, less than half of the previously used estimates~\cite{Plessel2003,Knebel2006,Steppke2010},
and obtain the Sommerfeld coefficient $\gamma_S = (1.65\pm 0.01)$\,J/(mol\,K\textsuperscript{2}), in good agreement with previous work \cite{Krellner2009}.


Fig.~\ref{fig:field_data}(a) shows the evolution of the $T_A$ anomaly with magnetic field $B_{ext}$ applied in the ab-plane in the range
0.0-21.1\,mT. The anomaly shifts to lower temperatures with increasing applied field, broadens, and possibly develops a structure (a split into a double peak is observed at 14.7\,mT). Measurements in fields in the range 35.9-69.7\,mT do not display any anomaly and their overall shape resembles a typical Schottky peak.

In all magnetic fields the data above $T_A$ (or down to the lowest temperatures at $B_{ext} > 35.9$\,mT, where the anomaly was not observed)
are well described by the Schottky model, assuming paramagnetic polarisation $\vec\mu_{e} \parallel \vec B_{ext}$, see Fig.~\ref{fig:field_data}(b).
Fixing the quadrupolar parameters for \textsuperscript{173}Yb at the zero field value,
we find approximately linear growth of $\mu_{e}$ with field up to $\approx 0.1\mu_{B}$ at $B_c = 60$\,mT,
with weaker increase at higher fields, Fig.\ref{fig:field_data}(c),
consistent with the magnetic susceptibility measurements~\cite{Friedemann2009,Brando2013,Steinke2017}.
More subtle effects, such as the temperature dependence of $\mu_{e}$, may improve the agreement between the data and the model.

\begin{figure}[t!]
    \centering
    \includegraphics{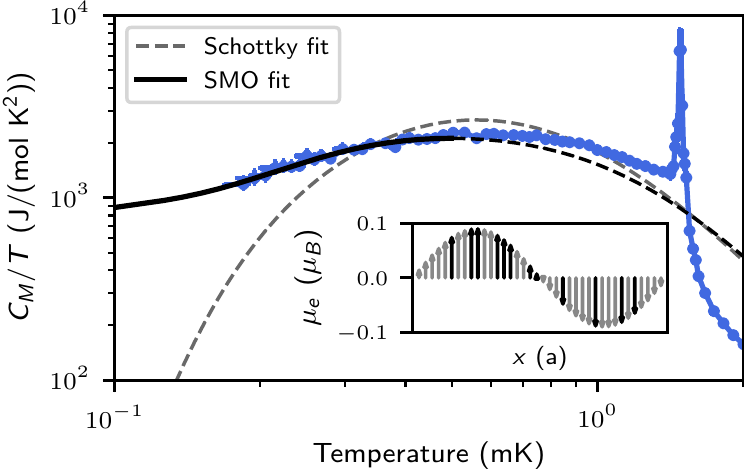}
    \caption{Zero field $C_M/T$ with the fit according to the SMO model below 0.5\,mK;
    the ``best fit'' according to spatially homogeneous Schottky model clearly disagrees with the data.
    The inset: electronic moments in SMO, black arrows represent randomly distributed Yb sites with active nuclei. 
    }
    \label{fig:SDW_fit}
\end{figure}

We now discuss the transition at $T_A$ and the data down to the lowest temperatures.
The entropy release below 10\,mK matches well the full nuclear entropy of \textsuperscript{171}Yb and \textsuperscript{173}Yb,
$S_{\mathrm{Yb}}=3.22$\,J/(mol\,K) in all magnetic fields, see Fig.~S1 in SI.
Under the conditions of our experiments the nuclear spins $^{29}$Si and $^{103}$Rh remain disordered and do not contribute to the heat capacity due to weak hyperfine constants for these elements.

Continuous warm-up measurements in zero magnetic field suggest, despite the sharpness of the heat capacity anomaly,
that the phase transition is continuous, see Fig.~S2 in SI. 
The majority of the Yb nuclear entropy is released below $T_A$, leaving only $0.06 S_{\mathrm{Yb}}$ for the transition region.
This points to gradual ordering of Yb nuclear spins in the hyperfine field produced by the electronic moments
and supports the picture of a nuclear-assisted electronic transition, developed later.

In zero magnetic field, the relatively slow decrease of the heat capacity with decreasing temperatures
cannot be accounted for by the Schottky model with uniform $\mu_{e}$, see Fig.~\ref{fig:SDW_fit}.
We therefore postulate a spatially modulated electronic order (SMO) state, with a sinusoidal distribution of the hyperfine field on the randomly distributed \textsuperscript{171}Yb and \textsuperscript{173}Yb nuclei (see inset in Fig.~\ref{fig:SDW_fit}) produced by the electronic moments
\begin{equation}
    \vec \mu_{e}(\vec{r})= \mu_A \vec{\hat{x}} \sin(\vec{r}\cdot\vec{q}), 
\label{eq:SMO}
\end{equation}
where $\vec{\hat{x}}$ is a unit vector, that we assume to lie in the easy ab plane. The heat capacity derived from the SMO model is insensitive to $\vec q$, as long as it is small or incommensurate, leaving a single free parameter, the maximum value of the modulated electronic moment $\mu_A$.
The fit to the data below 0.5\,mK yields $\mu_A = (0.093\pm0.001)\,\mu_B$.
It is noteworthy that this is comparable with the size of the moment induced by the critical field of the $T_N$ order, see Fig.~\ref{fig:field_data}(c).
Nuclear magnetic resonance is an established tool to confirm the existence of a SMO, or a spin density wave (SDW), in the absence of direct evidence from neutron scattering. Here we demonstrate that the heat capacity of nuclei responding to electronic order is another powerful probe of SMO,
albeit it does not allow us to determine $\vec q$ in Eq.~\eqref{eq:SMO}.

\begin{figure}[t!]
    \centering
    \includegraphics{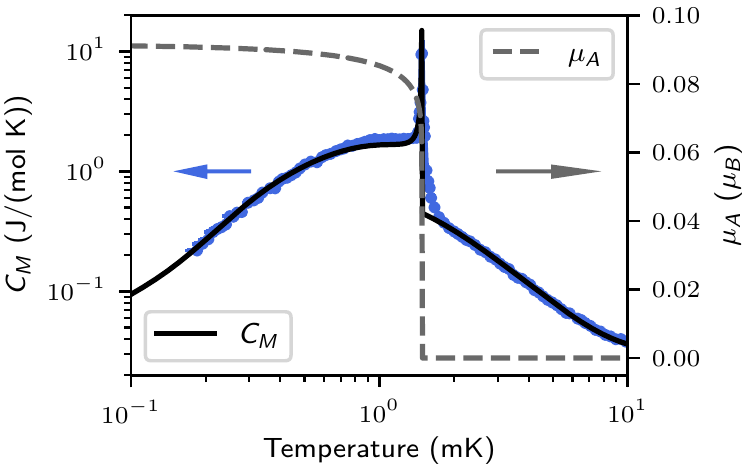}
    \caption{Model of the zero-field heat capacity using a simple ansatz for the temperature dependence of the order parameter, Eq.~\eqref{eq:Ansatz}.}
    \label{fig:Dynamic_C_ansatz_V}
\end{figure}

To account for the shape of the heat capacity anomaly, we make a simple ansatz for the temperature dependence of the order parameter
\begin{equation}
    \mu_A(T) = \mu_A(T=0)\left| 1 - T / T_A \right|^{\beta_c}.
\label{eq:Ansatz}
\end{equation}
For $\beta_c \approx 0.07$ the calculated heat capacity fits the zero-field data well across the whole temperature region, as shown in Fig.~\ref{fig:Dynamic_C_ansatz_V}. The smallness of the critical exponent $\beta_c$
and the resulting sharpness of the heat capacity peak demonstrate significant critical point fluctuations.

We now move to a simple mean-field model that captures the $T_A$ transition.
We argue that the mechanism behind this transition in the electronic magnetism is the hyperfine coupling of the static Yb electronic moments $\mu_{e}$ to the active Yb nuclei.
The key is to recognize the fragility of the AFM order that emerges at $T_N$.
The established dependence $\mu_{e} (B_{ext})$, Fig.~\ref{fig:field_data}(c), allows to determine the magnetic Helmholtz free energy
\begin{equation}
F_e(\mu_{e}) = \int_0^{\mu_{e}} B\,d\mu_{e}'
\label{eq:Penalty_energy}
\end{equation}
per Yb site.
At $B_{ext} = B_c = 70$\,mT, that suppresses the AFM order, we find $F_e(0.09\mu_B)\approx2$\,mK, much smaller than $T_N$.
The consequence of this small scale is that the energy cost of an increase in electronic moment can be overcome
by the reduction in the free energy of the nuclear spin system, polarised in the hyperfine field induced by these electronic moments.

\begin{figure}[t!]
    \centering
    \includegraphics{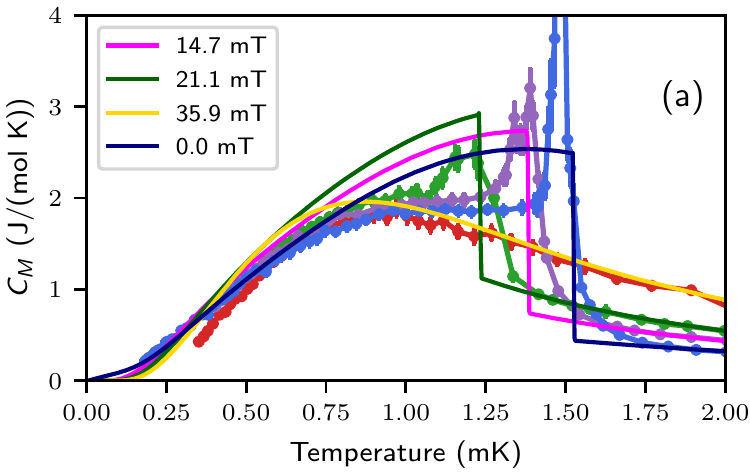}
    ~~~~~\includegraphics{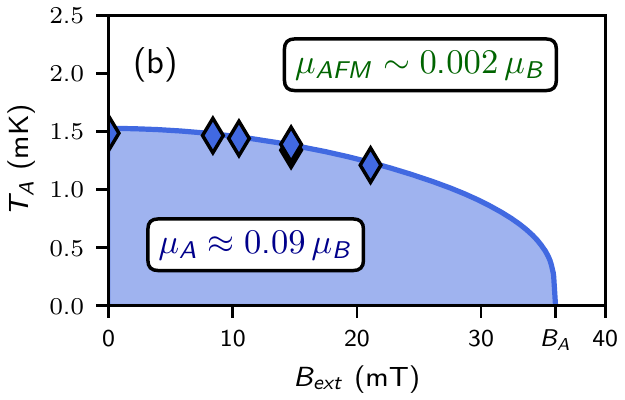}
    \caption{Comparison of the experimental data to the phenomenological mean-field model based on Eq.~\eqref{eq:Full_free_energy}.
    (a) Heat capacity. The sharpness of the anomaly is the only major feature not captured by the model.
    (b) Phase diagram of SMO in YbRh\textsubscript{2}Si\textsubscript{2}. Diamonds and straight line represent the experimental data and model respectively.
    }
    \label{fig:C_and_OP_favoured_SDW}
\end{figure}

To model the system through $T_A$ in an in-plane field applied along a unit vector $\vec{\hat{y}}$ we consider electronic moments
\begin{equation}
  \vec \mu_{e}(\vec r) = \mu_A \vec{\hat{x}} \sin (\vec r\cdot \vec q) + \mu_P\vec{\hat{y}},
  \label{eq:SMO:paro}
\end{equation}
with mutually orthogonal modulated $\mu_A$ and paramagnetic $\mu_P$ components, both in the ab plane. The model does not consider the pre-existing $T_N$ order due to its small moments. We write the Gibbs free energy as
\begin{align}
    &G(T, B_{ext}; \mu_A, \mu_P) = \sum_{\vec{r}} -k_BT\log Z\left(T, \vec{\mu}_{e}(\vec r)\right)
\label{eq:Full_free_energy} \\
    &\,\,+ N \left[\alpha\mu_A^2 + \beta\mu_P^2 + \gamma\mu_A^4 + \delta\mu_P^4 + \eta\mu_A^2\mu_P^2 - B_{ext}^{\phantom{2}}\mu_P^{\phantom{2}} \right],\notag
\end{align}
where the sum is over the active Yb sites with different values of the static electronic moment and size of nuclear spin $I$ and $N$ is the total number of Yb sites in the system.
At each site the nuclear partition function $Z$ is evaluated numerically from the Hamiltonian in Eq.~\eqref{eq:Hamilton}.
The phenomenological energy of the electronic order is expanded up to the fourth order in $\mu_A$ and $\mu_P$, retaining only the
terms allowed by symmetry.
The electronic order parameters $\mu_A(T, B_{ext})$ and $\mu_P(T, B_{ext})$ are found by minimising $G$, then the entropy and heat capacity are evaluated (see SI).


The model successfully captures most features of the experimental data, see Fig.~\ref{fig:C_and_OP_favoured_SDW}.
Here $\beta = 1/(2\chi) \approx 0.4$\,T/$\mu_B$ and $\delta = 0$ are fixed from the susceptibility measured at $T > T_A$~\cite{Brando2013};
$\alpha\approx0.026$\,T/$\mu_B$ is determined from $T_A$ at zero field; the non-essential 4-th order terms
$\gamma\approx0.2$\,T/$\mu_B^3$ and $\eta \approx 4$\,T/$\mu_B^3$ improve the agreement with the measured temperature
dependence of the heat capacity below $T_A$ and the suppression of $T_A$ with field.

At zero magnetic field the SMO is established via a second-order phase transition.
In-plane field additionally induces the paramagnetic component $\mu_P$, similar to that shown in Fig.~\ref{fig:field_data}(c),
and gradually suppresses $T_A$ and $\mu_A$, but the transition remains second-order up to the ultimate suppression of SMO at $B_A = 36$\,mT.
Both $T_A$ and $B_A$ are sensitive to the Yb isotopic composition.
The key features of this numerical model are illustrated by
a simple analytic $I=1/2$ model in SI.

We conclude by discussing the insight provided by these heat capacity measurements on superconductivity in YbRh\textsubscript{2}Si\textsubscript{2} \cite{Schuberth2016,Nguyen2021,Steinke2017,Smidman2018,Knappova2023,Levitin2022}. The heat capacity signature at $T_A$ is well aligned in temperature with an abrupt change in the shielding factor \cite{Schuberth2016,Steinke2017,Knappova2023} and sample electrical impedance \cite{Levitin2022}.
According to Ref.~\cite{Schuberth2016}, the superconducting transition occurs in the vicinity of $T_A$, however both the magnetic shielding \cite{Schuberth2016,Knappova2023} and electrical transport \cite{Nguyen2021,Levitin2022} exhibit sharp onset of superconductivity around 10\,mK. At $T_A$ the nuclear heat capacity dominates the heavy-electron term by at least 3 orders of magnitude.
Therefore any anomaly in the electronic heat capacity associated with a BCS-like superconducting transition at that temperature is undetectable in these measurements.
On the other hand at 6--12\,mK our data would reveal such a signature against the nuclear quadrupole background (see SI, Fig.~S9). Its absence points towards inhomogeneous superconductivity.

Refs.~\cite{Schuberth2016,Smidman2018} postulated competition between electronic magnetism and the superconductivity, and proposed that the latter is only established after the former is weakened at $T_A$. By contrast we find that below $T_A$, where the superconductivity is more robust, the electronic magnetism is simultaneously strengthened. Whether the AFM order established at $T_N$ is suppressed below $T_A$ remains an open question. The two magnetic orders may coexist and even share the same $\vec q$ (more in SI).

The observation of superconductivity both above and below $T_A$ opens an intriguing possibility of different superconducting order parameters in these regimes. The superconductivity and SMO may be intertwined, forming a pair density wave \cite{Liu2021,Fradkin2015} below $T_A$.

Our heat capacity method should be applied to the investigation of isotopically-enriched samples. This includes potential nuclear spin ordering in \textsuperscript{171}YbRh\textsubscript{2}Si\textsubscript{2} and \textsuperscript{173}YbRh\textsubscript{2}Si\textsubscript{2}, as well as heavy-fermion quantum criticality in \textsuperscript{174}YbRh\textsubscript{2}Si\textsubscript{2}, in the absence of nuclear magnetism.

We demonstrated how strong hyperfine interactions can give rise to a nuclear-assisted transition in the electronic magnetism at a temperature below the onset of superconductivity. This offers a fresh opportunity to further the understanding of the interplay between magnetism and superconductivity, accessible by new experimental techniques which extend studies of strongly correlated electron systems into the microkelvin regime.

\begin{acknowledgments}
This work was supported by the European Microkelvin Platform, supported by European Union's Horizon 2020 Research and Innovation programme under grant agreement no. 824109 and the Deutsche Forschungsgemeinschaft (DFG, German Research Foundation) through grants Nos. BR 4110/1-1, KR3831/4-1, and via the TRR 288, (422213477, project A03). We would like to thank Piers Coleman and S\'eamus Davis for helpful discussions.
\end{acknowledgments}

\onecolumngrid

\renewcommand{\thefigure}{S\arabic{figure}}
\renewcommand{\thetable}{S\arabic{table}}
\renewcommand{\theequation}{S\arabic{equation}}
\setcounter{figure}{0}
\setcounter{table}{0}
\setcounter{equation}{0}

\vskip2em

\centerline{\large\textbf{Supplemental Material}}
\vskip1em

\twocolumngrid 

\section{Entropy of $\mathbf{YbRh}$\textsubscript{2}$\mathbf{Si}$\textsubscript{2}}

The measured molar heat capacity of YbRh\textsubscript{2}Si\textsubscript{2} below 10\,mK comes almost exclusively from nuclear degrees of freedom of the half integer spin nuclei of Yb. All naturally occurring isotopes of Yb, Rh and Si are summarised in Table \ref{tab:Yb_isotopes}. The full entropy of spin $I$, regardless of the source of the level splitting (Zeeman, quadrupolar effect, etc.), is clearly $R\log(2I+1)$. The nuclear entropy is released at experimentally achievable temperatures only in sufficiently strong sources of Zeeman, or quadrupolar splitting. 
\begin{table}[b!]
    \caption[Isotopes of Yb, Rh and Si.]{Isotopes of Yb, Rh and Si, their nuclear spin $I$, gyromagnetic ratio $\gamma$, electric quadrupole $Q$ (from optical spectroscopy \cite{Weast1975}) and natural abundance.}
    \label{tab:Yb_isotopes}    \centering
    \begin{tabular}{ccccc}
    \hline\hline\rule{0pt}{1.1em}
        Isotope & $I$ & $\gamma$ [MHz/T] & $Q$ [10\textsuperscript{-28}m\textsuperscript{2}] & Abundance [\%] \\
    \hline\rule{0pt}{1.2em}
        \textsuperscript{174}Yb & 0 & - & - & 31.8 \\
        \textsuperscript{172}Yb & 0 & - & - & 21.8 \\
        \textsuperscript{173}Yb & 5/2 & -2.073 & 2.8 & 16.1 \\
        \textsuperscript{171}Yb & 1/2 & 7.52 & - & 14.3 \\
        \textsuperscript{176}Yb & 0 & - & - & 12.8 \\
        \textsuperscript{170}Yb & 0 & - & - & 3.0 \\
        \textsuperscript{168}Yb & 0 & - & - & 1.3 \\
    \hline\rule{0pt}{1.2em}
        \textsuperscript{103}Rh & 1/2 & -1.35 & - & 100 \\
    \hline\rule{0pt}{1.2em}
        \textsuperscript{28}Si & 0 & - & - & 92.2 \\
        \textsuperscript{29}Si & 1/2 & -8.465 & - & 4.7 \\
        \textsuperscript{30}Si & 0 & - & - & 3.1 \\
    \hline\hline
    \end{tabular}
\end{table}

\begin{figure}[t!]
    \centering
    \includegraphics{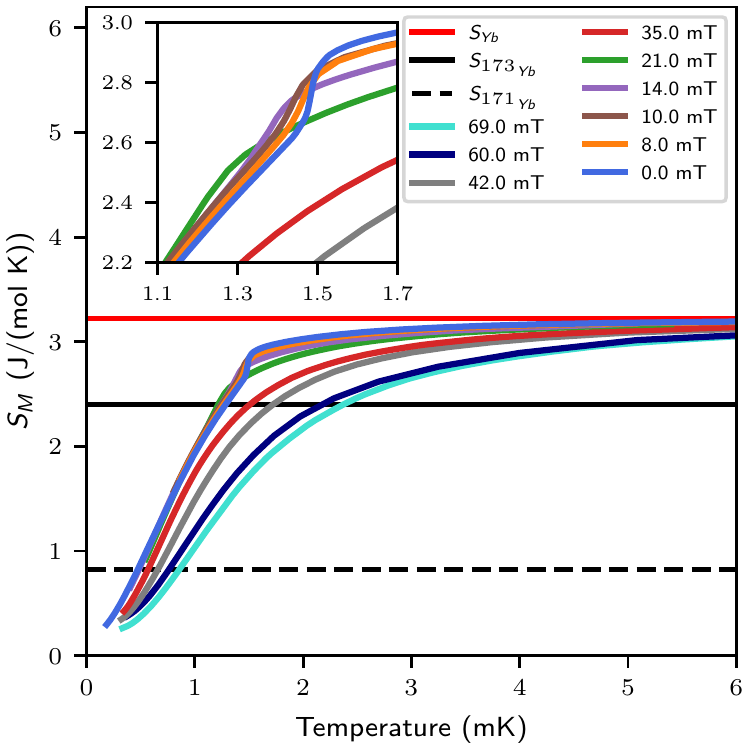}
    \caption{Entropy determined from the heat capacity measurements as a rolling integral. Electronic heat capacity subtracted, thus all curves converge to a horizontal line representing the full entropy of \textsuperscript{171}Yb and \textsuperscript{173}Yb nuclei. Inset shows the behaviour across the $T_A$ anomaly amplified.}
    \label{fig:entropy}
\end{figure}

The molar entropy at temperature $T$ can be calculated from the measured heat capacity as a rolling integral
\begin{equation}
    S_M(T) = \int_{0}^{T} \frac{C_M}{T'} dT'.
\label{eq:entropy_as_integral}
\end{equation}
The measurements however do not extend down to $T=0$. Another option is to integrate down from effectively infinite temperature, if one can subtract or neglect other contributions to the measured heat capacity. In case of YbRh\textsubscript{2}Si\textsubscript{2}, the electronic heat capacity is small and follows a simple linear temperature dependence of a (heavy) Fermi liquid. It is therefore easy to subtract. Fig.~\ref{fig:entropy} shows the nuclear entropy, calculated from the measured heat capacity, in all applied fields.

\begin{figure}[t!]
    \centering
    \includegraphics{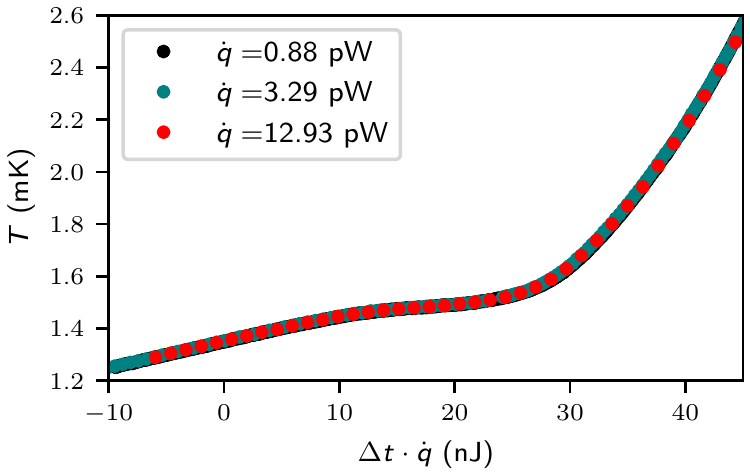}
    \caption{Continuous warm-up crossings of the $T_A$ anomaly in YbRh\textsubscript{2}Si\textsubscript{2} ($B_{ext} = 0.1$\,mT) at different speeds realized by applying different power in the heater. The legend specifies the overall power for the given warm-up curve, which consists of the applied power and a parasitic heat leak contribution, 80\,fW. The absence of a plateau and speed dependence speaks against a first order phase transition.}
    \label{fig:cont_warmup}
\end{figure}

\section{Nature of the $\bm{T_A}$ anomaly}

The steep, almost vertical, appearance of the $T_A$ anomaly in the entropy plot in Fig.~\ref{fig:entropy} raises the question whether the transition is first or second order. Continuous warm-up measurements are a very powerful method to distinguish the two cases. A plateau in $T(t)$ during a warm-up would be a direct proof of a first-order phase transition.
This was not observed, however the plateau can in principle be smeared by inhomogeneity.
The dependence on the warm-up rate is another signature of a first order phase transition. We have performed three continuous warm-ups across the anomaly with three different values of power applied in the heater. Their collapse is shown in Fig.~\ref{fig:cont_warmup} and supports the identification of $T_A$ as a continuous (second-order) phase transition.
The heat capacity extracted from these continuous measurements agrees quantitatively
with the results of the pulsed method.

\section{Nuclear Hamiltonian}

In this work we consider the Yb nuclear system to be governed by the single-spin Hamiltonian
\begin{equation}\tag{1}
    \begin{gathered}
    \widehat{H} = g\mu_N A_{hf} \vec{\widehat{I}}\cdot \vec{\mu}_{e} + \frac{e^2qQ}{4I(2I-1)}(3\widehat{I}_z^2 - I(I+1)),
    \end{gathered}
\end{equation}
where $g$, $\mu_N$, $Q$ are the nuclear g-factor, magneton and quadrupole moment and $eq$ represents the electric field gradient, which points along the c-axis.
Under the conditions of our experiment $B_{hf} \gg B_{ext}$,
therefore we do not include the Zeeman term $-g\mu_N \widehat{\vec I} \cdot \vec B_{ext}$ in the Hamiltonian.

The magnetic field applied in the ab-plane induces static electronic moments in the plane. We argue that the SMO established below $T_A$ also features electronic moments in the ab-plane. The energy levels of Eq.~\eqref{eq:Hamilton} do not
depend on the orientation of $\vec{\mu}_e \perp z$ in the ab-plane and for the numerical calculations
it is sufficient to assume $\vec{\mu}_e \parallel x$. In this case
$\widehat{\vec{I}} \cdot \vec{\mu}_e = \widehat I_x \mu_e$.
Due to symmetry this result holds even in case of SMO in field
where $\vec{\mu}_e(\vec r)$ are not aligned across the system,
since we do not consider direct interactions between nuclear spins.
For electronic moments along the c-axis, as has been proposed above $T_A$,
$\widehat{\vec{I}} \cdot \vec{\mu}_e = \widehat I_z \mu_e$.

In case of spin-$\frac{5}{2}$ \textsuperscript{173}Yb
\begin{equation}
    \widehat{I}_x = \frac{1}{2}
    \begin{pmatrix}
    0 & \sqrt{5} & 0 & 0 & 0 & 0 \\
    \sqrt{5} & 0 & \sqrt{8} & 0 & 0 & 0 \\
    0 & \sqrt{8} & 0 & \sqrt{9} & 0 & 0 \\
    0 & 0 & \sqrt{9} & 0 & \sqrt{8} & 0 \\
    0 & 0 & 0 & \sqrt{8} & 0 & \sqrt{5}\\
    0 & 0 & 0 & 0 & \sqrt{5} & 0\\
    \end{pmatrix},
\end{equation}
\begin{equation}
    \widehat{I}_z = \frac{1}{2} \textrm{diag} [5, 3, 1, -1, -3, -5].
\end{equation}

\begin{figure}[t!]
    \centering
    \includegraphics{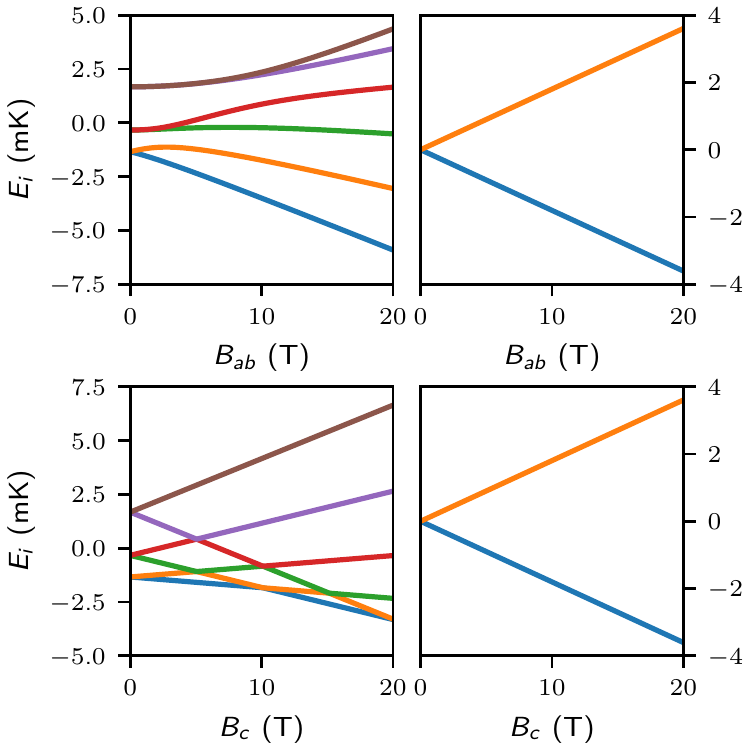}
    \caption{Energy levels of the single-spin Hamiltonian in Eq.~\eqref{eq:Hamilton} for \textsuperscript{173}Yb (left) and \textsuperscript{171}Yb (right). The large magnetic field is produced by the static electronic moment of Yb.}
    \label{fig:Single_ion_energy_levels}
\end{figure}
The spin-$\frac{1}{2}$ \textsuperscript{171}Yb isotope does not have a quadrupole moment,
hence the spin quantisation axis $z$ can be chosen to be parallel with $\vec{\mu}_e$, and
$\widehat{\vec{I}} \cdot \vec{\mu}_e = \widehat I_z \mu_e$, where
\begin{equation}
    \widehat{I}_z = \frac{1}{2} \textrm{diag} [1, -1].
\end{equation}
Generally, the \textsuperscript{173}Yb Hamiltonian must be diagonalised numerically. The energy levels of the two isotopes as a function of magnetic field applied along the c-axis and in the ab-plane are shown in Fig.~\ref{fig:Single_ion_energy_levels}.

The single-particle partition function of the nuclear system is
\begin{equation}
    Z \equiv \sum_{i=-I}^I e^{-E_i/k_BT}.
\end{equation}

\begin{figure}[t!]
    \centering
    \includegraphics{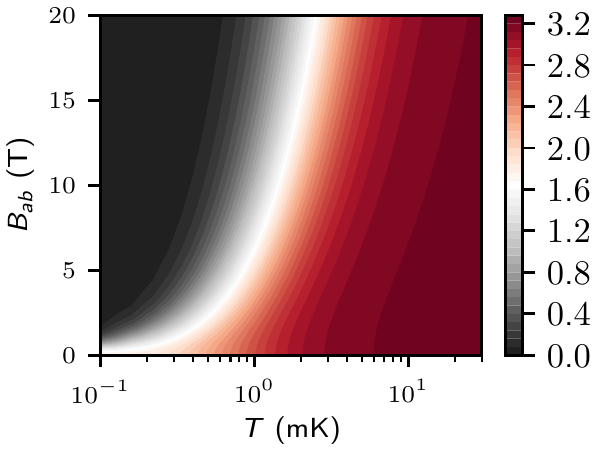}
    \caption{Contour plot of entropy of natural YbRh\textsubscript{2}Si\textsubscript{2} nuclear system as a function of temperature and magnetic field in the ab-plane. 
    }
    \label{fig:entropy_contour}
\end{figure}
In the case of temperature-independent (and spatially homogeneous) electronic moments, the \textit{Schottky model}, the heat capacity is calculated as
\begin{equation}
    \begin{gathered}
    C_M = \frac{R}{(k_B T)^2}\frac{\sum\limits_{i,j=-I}^I (E_i^2-E_iE_j)\,\exp[-(E_i+E_j)/k_B T]}{\sum\limits_{i,j=-I}^I \exp[-(E_i+E_j)/k_B T]}.
    \end{gathered}
\label{eq:Spec_heat_single_ion_H}
\end{equation}
Generally, the entropy of the nuclear system is
\begin{equation}
    S_M = -k_B\sum_{i=-I}^I p_i\log p_i,
\end{equation}
where $p_i = \exp[-E_i / k_B T] / Z$ is the probability of a particular micro-state. Then the heat capacity is calculated from the entropy as $C_M = T \partial S_M/\partial T$.

A colour plot of the YbRh\textsubscript{2}Si\textsubscript{2} nuclear system entropy as a function of temperature and field in the ab-plane is shown in Fig.~\ref{fig:entropy_contour}. An anstatz for the electronic order parameter $\vec\mu_e (T)$ defines a trajectory in this entropy landscape.

\section{Origins of the $\bm{T_A}$ transition, simple~spin-1/2~model}


Here we construct a simple model which considers nuclear spins $\frac{1}{2}$ coupled to
\emph{uniform} in-plane electronic moments $\mu_e$.
We only retain the leading order term $N \beta \mu_e^2$ in the energy cost of developing such moments for a system with $N$ Yb sites. Since such uniform moments develop in response to the in-plane magnetic field, $\beta = 1 / 2\chi$.
Out of the total $N$ Yb sites we assume $N_n$ to have non-zero nuclear spins.
The intensive variables of the problem are temperature and external magnetic field. The Gibbs free energy of our model is
\begin{align}
    G(T, B_{ext}; \mu_e) = &-N_n k_B T \log\left(2\cosh\left( \frac{g\mu_N A_{hf} \mu_e}{2k_BT} \right) \right)\notag\\
    &+ N\beta\mu_e^2.
\end{align}
To minimise the free energy with respect to the order parameter we solve
\begin{equation}
    \frac{\partial G}{\partial \mu_e} = -N_n\frac{g\mu_NA_{hf}}{2}\tanh\left( \frac{g\mu_NA_{hf}\mu_e}{2k_BT} \right) + 2N\beta\mu_e = 0.
\label{eq:B_minimum_simple2}
\end{equation}
This problem has an analytical solution~\cite{Cowan2021}
\begin{equation}
    \frac{T}{T_c} = \frac{2\mu_e/\mu_{e0}}{\log[(1+\mu_e/\mu_{e0})/(1-\mu_e/\mu_{e0})]},
\label{eq:simple_mod_solution}
\end{equation}
where
\begin{equation}
    \mu_{e0} = \frac{g\mu_N A_{hf}}{4\beta}\frac{N_n}{N}
\label{eq:moment_Tzero_simple_model}
\end{equation}
is the electronic moment at $T=0$ and the critical temperature
\begin{equation}
    T_c = \frac{g^2\mu_N^2A_{hf}^2}{8k_B\beta}\frac{N_n}{N}.
\label{eq:TA_simple_half_spin}
\end{equation}
In the vicinity of $T_c$ the magnetic moment is small and Eq.~\eqref{eq:simple_mod_solution} can be expanded
\begin{equation}
    1-\frac{T}{T_c} = \frac{1}{3}\left( \frac{\mu_e}{\mu_{e0}} \right)^2 + \frac{4}{45}\left(\frac{\mu_e}{\mu_{e0}} \right)^4 + \dots,
\end{equation}
giving a typical mean-field critical behaviour
\begin{equation}
    \frac{\mu_e}{\mu_{e0}} \approx \sqrt{3}\left( 1-\frac{T}{T_c} \right)^{1/2},
    \label{eq:simple_mod_solution:crit}
\end{equation}
Hence the model predicts the spontaneous growth of the magnetic moment at $T_c$ via a continuous phase transition.

Let us crudely approximate the Yb nuclear system in YbRh\textsubscript{2}Si\textsubscript{2} by $I = \frac{1}{2}$ spins,
by omitting the quadrupolar interaction and taking the Zeeman splitting of only the two furthest nuclear levels in case of \textsuperscript{173}Yb. Then based on Table~\ref{tab:Yb_isotopes} \textsuperscript{171}Yb and \textsuperscript{173}Yb have effective magnetic moments $0.49\mu_N$ and $0.68\mu_N$ respectively. Averaging by their individual abundances we obtain a magnetic moment $0.59\mu_N$ in 0.304 abundance.
Using $\beta=1/2\chi\approx0.4$\,T/$\mu_B$ from~\cite{Brando2013}
we obtain $T_c\approx270\,\mu$K and $\mu_{e0}=0.01\,\mu_B$.

The outcome of this illustrative and simple model is that even without the inclination of YbRh\textsubscript{2}Si\textsubscript{2} to develop SMO,
the presence of active nuclei, coupled to the electronic system by the hyperfine interactions, results in a growth of the static electronic moments in a ferromagnetic configuration.
An important result is the absence of a minimum density $N_n / N$ of nuclear spins required for such phase transition
to occur.

\section{Two perpendicular orders}

We continue in developing the analytical spin-$\frac{1}{2}$ model. In this section we consider
electronic order with both antiferromagnetic (A) and paramagnetic (P) components
\begin{equation}
  \vec \mu_{e}(\vec r) = \pm \mu_A \vec{\hat{x}} + \mu_P\vec{\hat{y}}.
  \label{eq:SMO:paro:simple}
\end{equation}
The two components are assumed to be perpendicular to each other;
the paramagnetic order is aligned with the applied field $B_{ext}$, if $B_{ext} \ne 0$.
Unlike SMO, in Eq.~\eqref{eq:SMO:paro:simple} the size of the antiferromagnetic moments, and therefore the total
electronic moment $\mu_e = \sqrt{\mu_A^2 + \mu_P^2}$ is the same at all Yb sites, enabling the analytical treatment.
We recognise that in a strongly-correlated electron system the costs of uniform and staggered electronic moments of the same magnitude are in general unequal. To the leading order in $\mu_A$ and $\mu_P$ we get the Gibbs free energy 
\begin{equation}
    \begin{aligned}
    &G(T, B_{ext}; \mu_ A, \mu_{P}) = \\
    &-N_nk_BT\log\left(2\cosh\left( \frac{g\mu_NA_{hf}\mu_e}{2k_BT} \right) \right)\\
    &+ N[\alpha\mu_{A}^2 + \beta\mu_{P}^2 - B_{ext}\mu_{P}].
    \end{aligned}
\label{eq:Gibbs_two_OPs}
\end{equation}
We find the order parameter values that minimise Eq.~\eqref{eq:Gibbs_two_OPs}
by solving 
\begin{equation}
    \begin{aligned}
    \frac{\partial G}{\partial \mu_{A}} = &-N_n\frac{g\mu_NA_{hf}}{2}\tanh\left( \frac{g\mu_NA_{hf}\mu_e}{2k_BT} \right)\frac{\mu_{A}}{\mu_e}\\
    &+ N[2\alpha\mu_{A}] = 0,\\
    \frac{\partial G}{\partial \mu_{P}} = &-N_n\frac{g\mu_NA_{hf}}{2}\tanh\left( \frac{g\mu_NA_{hf}\mu_e}{2k_BT} \right)\frac{\mu_{P}}{\mu_e}\\
    &+ N[2\beta\mu_{P} - B_{ext}] = 0,
    \end{aligned}
    \label{eq:Gibbs:conditions}
\end{equation}
and examining the sign of the determinant of the associated Hessian matrix.

In zero field the interplay between $\alpha$ and $\beta$ determines whether antiferromagnetic or uniform (ferromagnetic) order develops.
For $\alpha > \beta$ we recover the results of the previous section,
Eqs.~\eqref{eq:simple_mod_solution}-\eqref{eq:simple_mod_solution:crit}, with $\mu_P = \mu_e$, $\mu_A = 0$.

For $\alpha < \beta$ the continuous transition into AFM state occurs instead at
\begin{equation}
    T_A = \frac{g^2\mu_N^2A_{hf}^2}{8k_B\alpha}\frac{N_n}{N},
\end{equation}
higher than $T_c$ given by Eq.~\eqref{eq:TA_simple_half_spin}.
Here $\mu_P = 0$; the temperature dependence of $\mu_A = \mu_e$
is described by Eqs.~\eqref{eq:simple_mod_solution}-\eqref{eq:simple_mod_solution:crit} with $\beta$ and $T_c$ replaced by $\alpha$ and $T_A$.

We now examine the behaviour in magnetic fields at $T = 0$ assuming $\alpha < \beta$.
Here Eqs.~\eqref{eq:Gibbs:conditions} simplify to
\begin{subequations}\label{eq:T0:conditions}
\begin{eqnarray}
    \frac{\partial G}{\partial \mu_A} &=& N_n\frac{g\mu_NA_{hf}\mu_{A}}{2\mu_{e}} - 2N\alpha\mu_{A} = 0,
    \label{eq:T0:conditions:muP}
    \\
    \frac{\partial G}{\partial \mu_P} &=& N_n\frac{g\mu_NA_{hf}\mu_{P}}{2\mu_{e}} - N[2\beta\mu_{P} - B_{ext}] = 0.\qquad\,\,\,
    \label{eq:T0:conditions:muA}
\end{eqnarray}
\end{subequations}
If $\mu_A \neq 0$, from Eq.~\eqref{eq:T0:conditions:muA} we obtain
\begin{equation}
    \mu_{e}(T = 0) = \frac{g\mu_N A_{hf}}{4\alpha}\frac{N_n}{N},
\label{eq:middle_step}
\end{equation}
a modified version of Eq.~\eqref{eq:moment_Tzero_simple_model}. Interestingly we observe that the applied field does not influence the size of the electronic moment at $T=0$ until the complete suppression of $\mu_A$.
Substituting Eq.~\eqref{eq:middle_step} into Eq.~\eqref{eq:T0:conditions:muP} we obtain
the paramagnetic response in the presence of $\mu_A \neq 0$,
\begin{equation}
    \mu_{P} = \frac{B_{ext}}{2(\beta-\alpha)},
\label{eq:PM_OP_Tzero_double}
\end{equation}
comparable to $\mu_P = B_{ext} / 2\beta$ at $T \gg T_A$.
From Eq.~\eqref{eq:SMO:paro:simple} and \eqref{eq:PM_OP_Tzero_double} we obtain 
\begin{equation}
    \mu_{A} = \sqrt{\left[\frac{g\mu_N A_{hf}}{4\alpha}\frac{N_n}{N}\right]^2 - \left[\frac{B_{ext}}{2(\beta-\alpha)}\right]^2}.
\end{equation}
We observe that for small applied fields the suppression of $\mu_A$ with $B_{ext}$ is weak.
Upon increasing the field to
\begin{equation}
    B_A = \frac{g\mu_NA_{hf}(\beta-\alpha)}{2\alpha} \frac{N_n}{N}
\end{equation}
the size of AFM moments $\mu_A$ continuously drops to zero.

\begin{figure}[b!]
    \centering
    \includegraphics{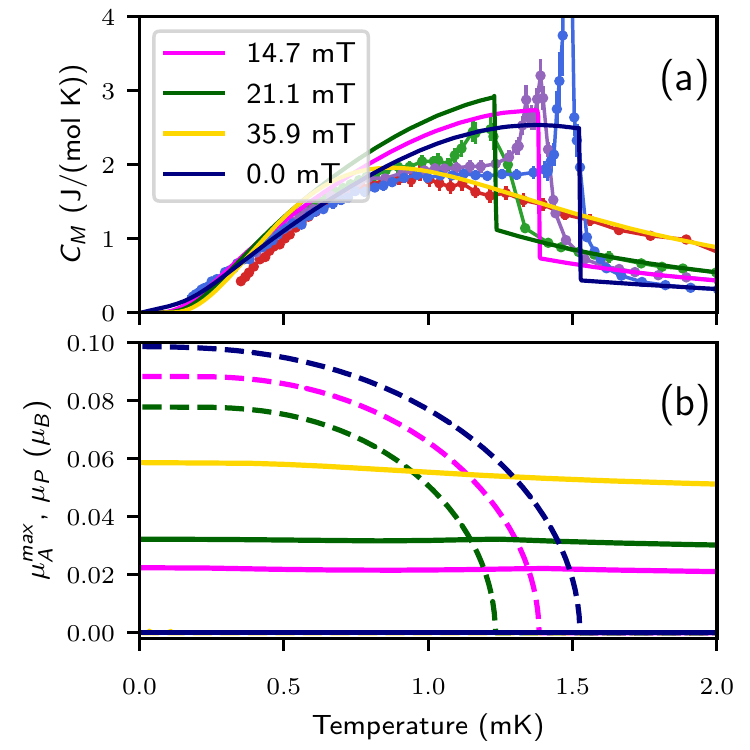}
    \caption{Comparison of the experimental data to the phenomenological mean-field model based on Eq.~\eqref{eq:Full_free_energy}.
    (a) Heat capacity. The sharpness of the anomaly is the only major feature not captured by the model.
    (b) Temperature dependence of the spatially modulated ($\mu_A$) and paramagnetic ($\mu_P$) orders
    shown with dashed and solid lines respectively.
    }
    \label{fig:C_and_OP_favoured_SDW_gamma}
\end{figure}

\begin{figure*}[t!]
    \centering
    \includegraphics{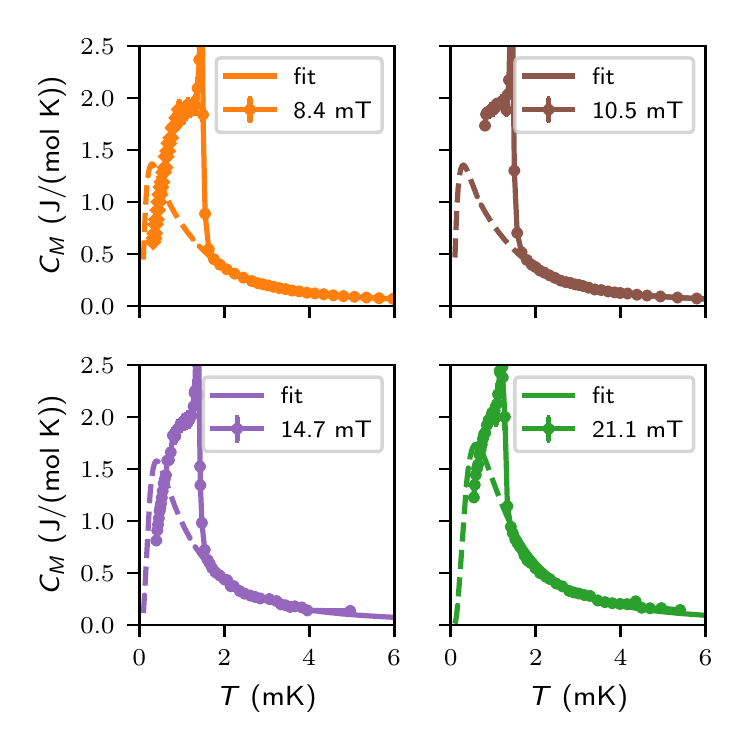}
    \includegraphics{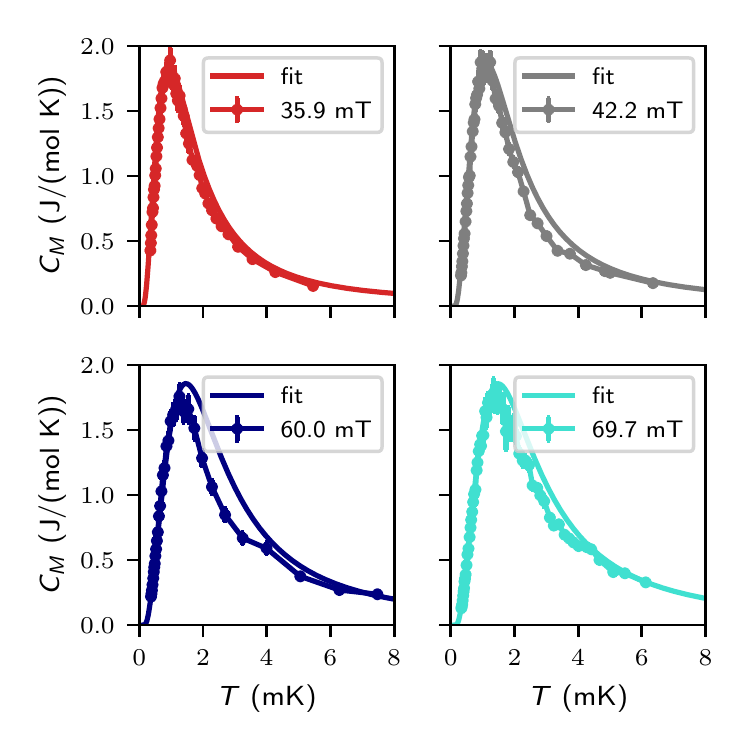}
    \caption{Schottky fits. At fields up to 21.1\,mT, $T_A$ anomaly was observed
    and only the data at $T > T_A$ were fitted.}
    \label{fig:Schottky_fits}
\end{figure*}

\section{Results of numerical modelling}

Here we provide some technical background of our full modelling of natural YbRh\textsubscript{2}Si\textsubscript{2}.
The fact that quadrupolar and hyperfine splitting of \textsuperscript{173}Yb nuclear spin are realized in mutually perpendicular axes complicates the problem and the results must be obtained numerically.
The full analysis was performed in Python 3. Eigenvalues of \textsuperscript{173}Yb Hamiltonian are found using \texttt{numpy} function \texttt{linalg.eig}.
The minimum of the Gibbs free energy is found using \texttt{scipy} function \texttt{optimize.minimize}.
Finding the minimum provides us with the order parameter(s).
Then the same approach is used to calculate the heat capacity as in the case of the ansatz for the order parameter.

The results of numerical modelling with two perpendicular orders $\mu_A$ and $\mu_P$ given by the full model
\begin{align}
    \vec \mu_{e}(\vec r) &= \mu_A \vec{\hat{x}} \sin (\vec r\cdot \vec q) + \mu_P\vec{\hat{y}},\tag{5}\\
    \tag{6}
    G(T, B_{ext}; \mu_A, \mu_P) &= \sum_{\vec{r}} -k_BT\log Z\left(T, \vec{\mu}_{e}(\vec r)\right)
\\
    \,\,+ N \big[\alpha\mu_A^2 + \beta\mu_P^2 &+ \gamma\mu_A^4 + \delta\mu_P^4 + \eta\mu_A^2\mu_P^2 - B_{ext}^{\phantom{2}}\mu_P^{\phantom{2}} \big],\notag
\end{align}
are shown in Fig.~\ref{fig:C_and_OP_favoured_SDW_gamma}, with the modelled heat capacity in (a) and the two order parameters in (b).

To strengthen the analogy with the simplest analytical model we study numerically the consequence of increasing $\alpha$ above $\beta = 1/2\chi \approx 0.4$\,T/$\mu_B$.
We obtain critical temperature $T_c=190\,\mu$K and the size of the ordered moment $\mu_P(T=0)=0.01\,\mu_B$, in good agreement with the analytical model.
The occurrence of a transition at the significantly higher temperature $T_A$ demonstrates that YbRh\textsubscript{2}Si\textsubscript{2} prefers a different order than spatially-homogeneous ferromagnet, which we identify to be the SMO.

\section*{Interplay between magnetic orders}

The direct effect on the nuclear heat capacity of the small staggered electronic moments $\mu_N \sim 0.002\mu_B$ above $T_A$ is negligible, and $\mu_N$ does not appear in our mean-field model, Eq.~(6).
As a result we ignore the interactions between $\mu_N$ and the order parameter components $\mu_A$ and $\mu_P$,
included in the model.
Let us rather suppose that the growth of $\mu_P$ in magnetic field or the development of $\mu_A$ at $T_A$ significantly affects $\mu_N$.
We recognise that the linear growth of the paramagnetic moments $\mu_P(B)$ with field
is observed experimentally up to $\mu_P \sim 0.1\mu_B$.
Thus for sufficiently small $\mu_P$ the electronic internal energy is well-described by the simple form $N\beta \mu_P^2$,
and any associated change in $\mu_N$ is implicitly taken into account.
We suggest that the same holds for the SMO, which is characterised by comparable electronic moments and energy cost,
therefore the behaviour of $\mu_A$ and $\mu_P$ with temperature and field is not significantly affected
by ignoring their interplay with $\mu_N$.
Nevertheless a detailed microscopic theory taking into account all components of the ordered electronic moments is desirable.

It it important to consider a scenario in which $\mu_N$, established at $T_N \gg T_A$, does not change significantly across the $T_A$ transition.
A plausible structure of coexisting antiferromagnetic orders involves mutually-orthogonal $\mu_N$ and $\mu_A$ with the same $\vec q$ and a $\pi/4$ phase shift,
\begin{equation}
\vec\mu_e(\vec r) = \mu_A \hat{\vec x} \sin (\vec r \cdot \vec q) + \mu_N \hat{\vec z} \cos (\vec r \cdot \vec q),
\end{equation}
where $\hat{\vec z}$ is a unit vector along the $c$ axis.
Here $\mu_A$ moments grow out of the nodes of the $\mu_N$, minimising the competition between the two orders
and making the SMO the preferred secondary order parameter.

\section{Schottky fit summary}

Data in all fields were fitted to the Schottky model, in the full temperature range for the data sets which do not display any anomaly, and above the $T_A$ anomaly, where it was observed, see Fig.~\ref{fig:Schottky_fits}.
The size of the electronic moment $\mu_e$ extracted as the single free parameter of these fits is shown in Fig.~2(c) in the Letter.

\begin{figure}[b!]
    \centering
    \includegraphics{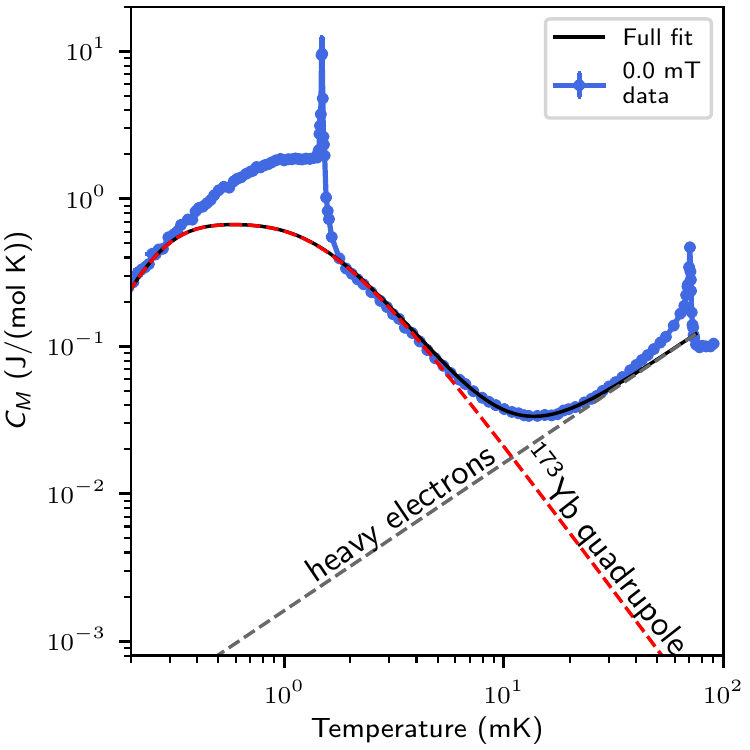}
    \caption{Nuclear and electronic contributions to the heat capacity at zero field.}
    \label{fig:Nuclear_electronic_contributions}
\end{figure}

\begin{figure}[t!]
    \centering
    \includegraphics{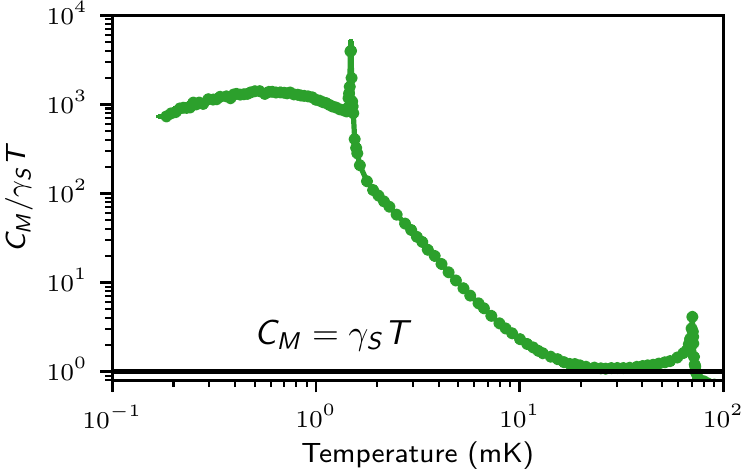}
    \caption{Zero-field heat capacity normalised by the heavy-electron term $\gamma_S T$, where $\gamma_S=1.65$\,J/(mol K\textsuperscript{2}).}
    \label{fig:CovGammaT}
    \vskip1em
    \centering
    \includegraphics{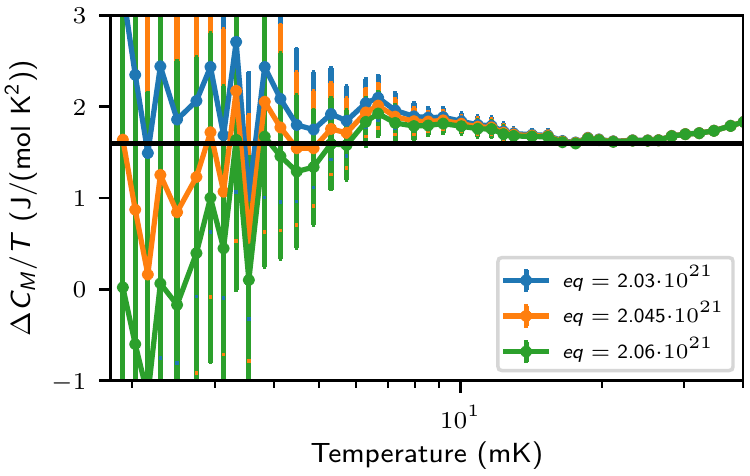}
    \caption{Heavy-electron heat capacity normalised by temperature,
    $\Delta C_M / T = (C_M - C_M^{\mathrm{nuclear}}) / T$,
    with the nuclear quadrupolar heat capacity $C_M^{\mathrm{nuclear}}$
    subtracted using three values of CEF gradient $eq$.
    The black horizontal line corresponds to Sommerfeld coefficient
    $\gamma_S = 1.65$\,J/(mol K\textsuperscript{2}).
    The heavy-fermion model agrees with the data up to approximately 30\,mK (vertical dashed line), 
    the tail of the N\'eel transition anomaly
    becomes significant at higher temperatures.}
    \label{fig:Chf}
\end{figure}
\section{Zero field heat capacity: search for signatures of superconductivity}

The fit in Fig.~1 in the Letter is a sum of nuclear and heavy-electron terms, where the nuclear part is dominated by the quadrupolar splitting of \textsuperscript{173}Yb nuclei. These terms are shown individually in Fig.~\ref{fig:Nuclear_electronic_contributions}, demonstrating that the nuclei and electrons dominate heat capacity below and above 10\,mK respectively,
with the relative strength of the electronic term illustrated in Fig.~\ref{fig:CovGammaT}.
While a BCS-like anomaly near $T_A$ would be invisible
on the background of the magnetic transition into SMO,
it could be resolved over the quadrupolar background above 6\,mK.

In search for signatures of superconductivity in this temperature range we subtract the quadrupolar nuclear heat capacity using three plausible values of the electric field gradient $eq$, see Fig.~\ref{fig:Chf}.
In this $C_M/T$ vs $T$ plot the heavy-electron heat capacity
is manifested by a horizontal line.
The data are consistent with this model up to
approximately 30\,mK where the tail
of the N\'eel transition anomaly starts to play a role.

No sharp BCS-type anomaly is observed, however there is a broad maximum centered around 7\,mK, on the border of statistical significance.
This may be a signature of inhomogeneous superconductivity.


\begin{thebibliography}{55}%
\makeatletter
\providecommand \@ifxundefined [1]{%
 \@ifx{#1\undefined}
}%
\providecommand \@ifnum [1]{%
 \ifnum #1\expandafter \@firstoftwo
 \else \expandafter \@secondoftwo
 \fi
}%
\providecommand \@ifx [1]{%
 \ifx #1\expandafter \@firstoftwo
 \else \expandafter \@secondoftwo
 \fi
}%
\providecommand \natexlab [1]{#1}%
\providecommand \enquote  [1]{``#1''}%
\providecommand \bibnamefont  [1]{#1}%
\providecommand \bibfnamefont [1]{#1}%
\providecommand \citenamefont [1]{#1}%
\providecommand \href@noop [0]{\@secondoftwo}%
\providecommand \href [0]{\begingroup \@sanitize@url \@href}%
\providecommand \@href[1]{\@@startlink{#1}\@@href}%
\providecommand \@@href[1]{\endgroup#1\@@endlink}%
\providecommand \@sanitize@url [0]{\catcode `\\12\catcode `\$12\catcode
  `\&12\catcode `\#12\catcode `\^12\catcode `\_12\catcode `\%12\relax}%
\providecommand \@@startlink[1]{}%
\providecommand \@@endlink[0]{}%
\providecommand \url  [0]{\begingroup\@sanitize@url \@url }%
\providecommand \@url [1]{\endgroup\@href {#1}{\urlprefix }}%
\providecommand \urlprefix  [0]{URL }%
\providecommand \Eprint [0]{\href }%
\providecommand \doibase [0]{https://doi.org/}%
\providecommand \selectlanguage [0]{\@gobble}%
\providecommand \bibinfo  [0]{\@secondoftwo}%
\providecommand \bibfield  [0]{\@secondoftwo}%
\providecommand \translation [1]{[#1]}%
\providecommand \BibitemOpen [0]{}%
\providecommand \bibitemStop [0]{}%
\providecommand \bibitemNoStop [0]{.\EOS\space}%
\providecommand \EOS [0]{\spacefactor3000\relax}%
\providecommand \BibitemShut  [1]{\csname bibitem#1\endcsname}%
\let\auto@bib@innerbib\@empty
\bibitem [{\citenamefont {Schuberth}\ \emph {et~al.}(2016)\citenamefont
  {Schuberth}, \citenamefont {Tippmann}, \citenamefont {Steinke}, \citenamefont
  {Lausberg}, \citenamefont {Steppke}, \citenamefont {Brando}, \citenamefont
  {Krellner}, \citenamefont {Geibel}, \citenamefont {Yu}, \citenamefont {Si},\
  and\ \citenamefont {Steglich}}]{Schuberth2016}%
  \BibitemOpen
  \bibfield  {author} {\bibinfo {author} {\bibfnamefont {E.}~\bibnamefont
  {Schuberth}}, \bibinfo {author} {\bibfnamefont {M.}~\bibnamefont {Tippmann}},
  \bibinfo {author} {\bibfnamefont {L.}~\bibnamefont {Steinke}}, \bibinfo
  {author} {\bibfnamefont {S.}~\bibnamefont {Lausberg}}, \bibinfo {author}
  {\bibfnamefont {A.}~\bibnamefont {Steppke}}, \bibinfo {author} {\bibfnamefont
  {M.}~\bibnamefont {Brando}}, \bibinfo {author} {\bibfnamefont
  {C.}~\bibnamefont {Krellner}}, \bibinfo {author} {\bibfnamefont
  {C.}~\bibnamefont {Geibel}}, \bibinfo {author} {\bibfnamefont
  {R.}~\bibnamefont {Yu}}, \bibinfo {author} {\bibfnamefont {Q.}~\bibnamefont
  {Si}},\ and\ \bibinfo {author} {\bibfnamefont {F.}~\bibnamefont {Steglich}},\
  }\bibfield  {title} {\bibinfo {title} {Emergence of superconductivity in the
  canonical heavy{\textendash}electron metal {YbRh}$_2${Si}$_2$},\ }\href
  {https://doi.org/10.1126/science.aaa9733} {\bibfield  {journal} {\bibinfo
  {journal} {Science}\ }\textbf {\bibinfo {volume} {351}},\ \bibinfo {pages}
  {485} (\bibinfo {year} {2016})}\BibitemShut {NoStop}%
\bibitem [{\citenamefont {Friedemann}\ \emph {et~al.}(2010)\citenamefont
  {Friedemann}, \citenamefont {Wirth}, \citenamefont {Oeschler}, \citenamefont
  {Krellner}, \citenamefont {Geibel}, \citenamefont {Steglich}, \citenamefont
  {MaQuilon}, \citenamefont {Fisk}, \citenamefont {Paschen},\ and\
  \citenamefont {Zwicknagl}}]{Friedemann2010b}%
  \BibitemOpen
  \bibfield  {author} {\bibinfo {author} {\bibfnamefont {S.}~\bibnamefont
  {Friedemann}}, \bibinfo {author} {\bibfnamefont {S.}~\bibnamefont {Wirth}},
  \bibinfo {author} {\bibfnamefont {N.}~\bibnamefont {Oeschler}}, \bibinfo
  {author} {\bibfnamefont {C.}~\bibnamefont {Krellner}}, \bibinfo {author}
  {\bibfnamefont {C.}~\bibnamefont {Geibel}}, \bibinfo {author} {\bibfnamefont
  {F.}~\bibnamefont {Steglich}}, \bibinfo {author} {\bibfnamefont
  {S.}~\bibnamefont {MaQuilon}}, \bibinfo {author} {\bibfnamefont
  {Z.}~\bibnamefont {Fisk}}, \bibinfo {author} {\bibfnamefont {S.}~\bibnamefont
  {Paschen}},\ and\ \bibinfo {author} {\bibfnamefont {G.}~\bibnamefont
  {Zwicknagl}},\ }\bibfield  {title} {\bibinfo {title} {Hall effect
  measurements and electronic structure calculations on {YbRh$_2$Si$_2$} and
  its reference compounds {LuRh$_2$Si$_2$} and {YbIr$_2$Si$_2$}},\ }\href
  {https://doi.org/10.1103/physrevb.82.035103} {\bibfield  {journal} {\bibinfo
  {journal} {Physical Review B}\ }\textbf {\bibinfo {volume} {82}},\ \bibinfo
  {pages} {035103} (\bibinfo {year} {2010})}\BibitemShut {NoStop}%
\bibitem [{\citenamefont {Kummer}\ \emph {et~al.}(2015)\citenamefont {Kummer},
  \citenamefont {Patil}, \citenamefont {Chikina}, \citenamefont
  {G{\"{u}}ttler}, \citenamefont {H{\"{o}}ppner}, \citenamefont {Generalov},
  \citenamefont {Danzenb{\"{a}}cher}, \citenamefont {Seiro}, \citenamefont
  {Hannaske}, \citenamefont {Krellner}, \citenamefont {Kucherenko},
  \citenamefont {Shi}, \citenamefont {Radovic}, \citenamefont {Rienks},
  \citenamefont {Zwicknagl}, \citenamefont {Matho}, \citenamefont {Allen},
  \citenamefont {Laubschat}, \citenamefont {Geibel},\ and\ \citenamefont
  {Vyalikh}}]{Kummer2015}%
  \BibitemOpen
  \bibfield  {author} {\bibinfo {author} {\bibfnamefont {K.}~\bibnamefont
  {Kummer}}, \bibinfo {author} {\bibfnamefont {S.}~\bibnamefont {Patil}},
  \bibinfo {author} {\bibfnamefont {A.}~\bibnamefont {Chikina}}, \bibinfo
  {author} {\bibfnamefont {M.}~\bibnamefont {G{\"{u}}ttler}}, \bibinfo {author}
  {\bibfnamefont {M.}~\bibnamefont {H{\"{o}}ppner}}, \bibinfo {author}
  {\bibfnamefont {A.}~\bibnamefont {Generalov}}, \bibinfo {author}
  {\bibfnamefont {S.}~\bibnamefont {Danzenb{\"{a}}cher}}, \bibinfo {author}
  {\bibfnamefont {S.}~\bibnamefont {Seiro}}, \bibinfo {author} {\bibfnamefont
  {A.}~\bibnamefont {Hannaske}}, \bibinfo {author} {\bibfnamefont
  {C.}~\bibnamefont {Krellner}}, \bibinfo {author} {\bibfnamefont
  {Y.}~\bibnamefont {Kucherenko}}, \bibinfo {author} {\bibfnamefont
  {M.}~\bibnamefont {Shi}}, \bibinfo {author} {\bibfnamefont {M.}~\bibnamefont
  {Radovic}}, \bibinfo {author} {\bibfnamefont {E.}~\bibnamefont {Rienks}},
  \bibinfo {author} {\bibfnamefont {G.}~\bibnamefont {Zwicknagl}}, \bibinfo
  {author} {\bibfnamefont {K.}~\bibnamefont {Matho}}, \bibinfo {author}
  {\bibfnamefont {J.~W.}\ \bibnamefont {Allen}}, \bibinfo {author}
  {\bibfnamefont {C.}~\bibnamefont {Laubschat}}, \bibinfo {author}
  {\bibfnamefont {C.}~\bibnamefont {Geibel}},\ and\ \bibinfo {author}
  {\bibfnamefont {D.~V.}\ \bibnamefont {Vyalikh}},\ }\bibfield  {title}
  {\bibinfo {title} {Temperature-independent {Fermi} surface in the {Kondo}
  lattice {YbRh\textsubscript{2}Si\textsubscript{2}}},\ }\href
  {https://doi.org/10.1103/physrevx.5.011028} {\bibfield  {journal} {\bibinfo
  {journal} {Physical Review X}\ }\textbf {\bibinfo {volume} {5}},\ \bibinfo
  {pages} {011028} (\bibinfo {year} {2015})}\BibitemShut {NoStop}%
\bibitem [{\citenamefont {Zwicknagl}(2016)}]{Zwicknagl2016}%
  \BibitemOpen
  \bibfield  {author} {\bibinfo {author} {\bibfnamefont {G.}~\bibnamefont
  {Zwicknagl}},\ }\bibfield  {title} {\bibinfo {title} {The utility of band
  theory in strongly correlated electron systems},\ }\href
  {https://doi.org/10.1088/0034-4885/79/12/124501} {\bibfield  {journal}
  {\bibinfo  {journal} {Reports on Progress in Physics}\ }\textbf {\bibinfo
  {volume} {79}},\ \bibinfo {pages} {124501} (\bibinfo {year}
  {2016})}\BibitemShut {NoStop}%
\bibitem [{\citenamefont {Li}\ \emph {et~al.}(2019)\citenamefont {Li},
  \citenamefont {Wang}, \citenamefont {Xu}, \citenamefont {Xie},\ and\
  \citenamefont {Yang}}]{Li2019a}%
  \BibitemOpen
  \bibfield  {author} {\bibinfo {author} {\bibfnamefont {Y.}~\bibnamefont
  {Li}}, \bibinfo {author} {\bibfnamefont {Q.}~\bibnamefont {Wang}}, \bibinfo
  {author} {\bibfnamefont {Y.}~\bibnamefont {Xu}}, \bibinfo {author}
  {\bibfnamefont {W.}~\bibnamefont {Xie}},\ and\ \bibinfo {author}
  {\bibfnamefont {Y.~F.}\ \bibnamefont {Yang}},\ }\bibfield  {title} {\bibinfo
  {title} {Nearly degenerate $p_x+ip_y$ and $d_{x^2-y^2}$ pairing symmetry in
  the heavy fermion superconductor {YbRh$_2$Si$_2$}},\ }\href
  {https://doi.org/10.1103/physrevb.100.085132} {\bibfield  {journal} {\bibinfo
   {journal} {Physical Review B}\ }\textbf {\bibinfo {volume} {100}},\ \bibinfo
  {pages} {085132} (\bibinfo {year} {2019})}\BibitemShut {NoStop}%
\bibitem [{\citenamefont {G{\"{u}}ttler}\ \emph {et~al.}(2021)\citenamefont
  {G{\"{u}}ttler}, \citenamefont {Kummer}, \citenamefont {Kliemt},
  \citenamefont {Krellner}, \citenamefont {Seiro}, \citenamefont {Geibel},
  \citenamefont {Laubschat}, \citenamefont {Kubo}, \citenamefont {Sakurai},
  \citenamefont {Vyalikh},\ and\ \citenamefont {Koizumi}}]{Guettler2021}%
  \BibitemOpen
  \bibfield  {author} {\bibinfo {author} {\bibfnamefont {M.}~\bibnamefont
  {G{\"{u}}ttler}}, \bibinfo {author} {\bibfnamefont {K.}~\bibnamefont
  {Kummer}}, \bibinfo {author} {\bibfnamefont {K.}~\bibnamefont {Kliemt}},
  \bibinfo {author} {\bibfnamefont {C.}~\bibnamefont {Krellner}}, \bibinfo
  {author} {\bibfnamefont {S.}~\bibnamefont {Seiro}}, \bibinfo {author}
  {\bibfnamefont {C.}~\bibnamefont {Geibel}}, \bibinfo {author} {\bibfnamefont
  {C.}~\bibnamefont {Laubschat}}, \bibinfo {author} {\bibfnamefont
  {Y.}~\bibnamefont {Kubo}}, \bibinfo {author} {\bibfnamefont {Y.}~\bibnamefont
  {Sakurai}}, \bibinfo {author} {\bibfnamefont {D.~V.}\ \bibnamefont
  {Vyalikh}},\ and\ \bibinfo {author} {\bibfnamefont {A.}~\bibnamefont
  {Koizumi}},\ }\bibfield  {title} {\bibinfo {title} {Visualizing the {Kondo}
  lattice crossover in {YbRh\textsubscript{2}Si\textsubscript{2}} with
  {Compton} scattering},\ }\href {https://doi.org/10.1103/physrevb.103.115126}
  {\bibfield  {journal} {\bibinfo  {journal} {Physical Review B}\ }\textbf
  {\bibinfo {volume} {103}},\ \bibinfo {pages} {115126} (\bibinfo {year}
  {2021})}\BibitemShut {NoStop}%
\bibitem [{\citenamefont {Ishida}\ \emph {et~al.}(2003)\citenamefont {Ishida},
  \citenamefont {MacLaughlin}, \citenamefont {Young}, \citenamefont {Okamoto},
  \citenamefont {Kawasaki}, \citenamefont {Kitaoka}, \citenamefont
  {Nieuwenhuys}, \citenamefont {Heffner}, \citenamefont {Bernal}, \citenamefont
  {Higemoto}, \citenamefont {Koda}, \citenamefont {Kadono}, \citenamefont
  {Trovarelli}, \citenamefont {Geibel},\ and\ \citenamefont
  {Steglich}}]{Ishida2003}%
  \BibitemOpen
  \bibfield  {author} {\bibinfo {author} {\bibfnamefont {K.}~\bibnamefont
  {Ishida}}, \bibinfo {author} {\bibfnamefont {D.~E.}\ \bibnamefont
  {MacLaughlin}}, \bibinfo {author} {\bibfnamefont {B.~L.}\ \bibnamefont
  {Young}}, \bibinfo {author} {\bibfnamefont {K.}~\bibnamefont {Okamoto}},
  \bibinfo {author} {\bibfnamefont {Y.}~\bibnamefont {Kawasaki}}, \bibinfo
  {author} {\bibfnamefont {Y.}~\bibnamefont {Kitaoka}}, \bibinfo {author}
  {\bibfnamefont {G.~J.}\ \bibnamefont {Nieuwenhuys}}, \bibinfo {author}
  {\bibfnamefont {R.~H.}\ \bibnamefont {Heffner}}, \bibinfo {author}
  {\bibfnamefont {O.~O.}\ \bibnamefont {Bernal}}, \bibinfo {author}
  {\bibfnamefont {W.}~\bibnamefont {Higemoto}}, \bibinfo {author}
  {\bibfnamefont {A.}~\bibnamefont {Koda}}, \bibinfo {author} {\bibfnamefont
  {R.}~\bibnamefont {Kadono}}, \bibinfo {author} {\bibfnamefont
  {O.}~\bibnamefont {Trovarelli}}, \bibinfo {author} {\bibfnamefont
  {C.}~\bibnamefont {Geibel}},\ and\ \bibinfo {author} {\bibfnamefont
  {F.}~\bibnamefont {Steglich}},\ }\bibfield  {title} {\bibinfo {title}
  {Low{\textendash}temperature magnetic order and spin dynamics in
  {YbRh}$_2${Si}$_2$},\ }\href {https://doi.org/10.1103/physrevb.68.184401}
  {\bibfield  {journal} {\bibinfo  {journal} {Physical Review B}\ }\textbf
  {\bibinfo {volume} {68}},\ \bibinfo {pages} {184401} (\bibinfo {year}
  {2003})}\BibitemShut {NoStop}%
\bibitem [{\citenamefont {Gegenwart}\ \emph {et~al.}(2002)\citenamefont
  {Gegenwart}, \citenamefont {Custers}, \citenamefont {Geibel}, \citenamefont
  {Neumaier}, \citenamefont {Tayama}, \citenamefont {Tenya}, \citenamefont
  {Trovarelli},\ and\ \citenamefont {Steglich}}]{Gegenwart2002}%
  \BibitemOpen
  \bibfield  {author} {\bibinfo {author} {\bibfnamefont {P.}~\bibnamefont
  {Gegenwart}}, \bibinfo {author} {\bibfnamefont {J.}~\bibnamefont {Custers}},
  \bibinfo {author} {\bibfnamefont {C.}~\bibnamefont {Geibel}}, \bibinfo
  {author} {\bibfnamefont {K.}~\bibnamefont {Neumaier}}, \bibinfo {author}
  {\bibfnamefont {T.}~\bibnamefont {Tayama}}, \bibinfo {author} {\bibfnamefont
  {K.}~\bibnamefont {Tenya}}, \bibinfo {author} {\bibfnamefont
  {O.}~\bibnamefont {Trovarelli}},\ and\ \bibinfo {author} {\bibfnamefont
  {F.}~\bibnamefont {Steglich}},\ }\bibfield  {title} {\bibinfo {title}
  {Magnetic-field induced quantum critical point in {YbRh$_2$Si$_2$}},\ }\href
  {https://doi.org/10.1103/physrevlett.89.056402} {\bibfield  {journal}
  {\bibinfo  {journal} {Physical Review Letters}\ }\textbf {\bibinfo {volume}
  {89}},\ \bibinfo {pages} {056402} (\bibinfo {year} {2002})}\BibitemShut
  {NoStop}%
\bibitem [{\citenamefont {Hamann}\ \emph {et~al.}(2019)\citenamefont {Hamann},
  \citenamefont {Zhang}, \citenamefont {Jang}, \citenamefont {Hannaske},
  \citenamefont {Steinke}, \citenamefont {Lausberg}, \citenamefont {Pedrero},
  \citenamefont {Klingner}, \citenamefont {Baenitz}, \citenamefont {Steglich},
  \citenamefont {Krellner}, \citenamefont {Geibel},\ and\ \citenamefont
  {Brando}}]{Hamann2019}%
  \BibitemOpen
  \bibfield  {author} {\bibinfo {author} {\bibfnamefont {S.}~\bibnamefont
  {Hamann}}, \bibinfo {author} {\bibfnamefont {J.}~\bibnamefont {Zhang}},
  \bibinfo {author} {\bibfnamefont {D.}~\bibnamefont {Jang}}, \bibinfo {author}
  {\bibfnamefont {A.}~\bibnamefont {Hannaske}}, \bibinfo {author}
  {\bibfnamefont {L.}~\bibnamefont {Steinke}}, \bibinfo {author} {\bibfnamefont
  {S.}~\bibnamefont {Lausberg}}, \bibinfo {author} {\bibfnamefont
  {L.}~\bibnamefont {Pedrero}}, \bibinfo {author} {\bibfnamefont
  {C.}~\bibnamefont {Klingner}}, \bibinfo {author} {\bibfnamefont
  {M.}~\bibnamefont {Baenitz}}, \bibinfo {author} {\bibfnamefont
  {F.}~\bibnamefont {Steglich}}, \bibinfo {author} {\bibfnamefont
  {C.}~\bibnamefont {Krellner}}, \bibinfo {author} {\bibfnamefont
  {C.}~\bibnamefont {Geibel}},\ and\ \bibinfo {author} {\bibfnamefont
  {M.}~\bibnamefont {Brando}},\ }\bibfield  {title} {\bibinfo {title}
  {Evolution from ferromagnetism to antiferromagnetism in
  {Yb(Rh\textsubscript{1-x}Co\textsubscript{x})Si\textsubscript{2}}},\ }\href
  {https://doi.org/10.1103/physrevlett.122.077202} {\bibfield  {journal}
  {\bibinfo  {journal} {Physical Review Letters}\ }\textbf {\bibinfo {volume}
  {122}},\ \bibinfo {pages} {077202} (\bibinfo {year} {2019})}\BibitemShut
  {NoStop}%
\bibitem [{\citenamefont {Stock}\ \emph {et~al.}(2012)\citenamefont {Stock},
  \citenamefont {Broholm}, \citenamefont {Demmel}, \citenamefont {{Van Duijn}},
  \citenamefont {Taylor}, \citenamefont {Kang}, \citenamefont {Hu},\ and\
  \citenamefont {Petrovic}}]{Stock2012}%
  \BibitemOpen
  \bibfield  {author} {\bibinfo {author} {\bibfnamefont {C.}~\bibnamefont
  {Stock}}, \bibinfo {author} {\bibfnamefont {C.}~\bibnamefont {Broholm}},
  \bibinfo {author} {\bibfnamefont {F.}~\bibnamefont {Demmel}}, \bibinfo
  {author} {\bibfnamefont {J.}~\bibnamefont {{Van Duijn}}}, \bibinfo {author}
  {\bibfnamefont {J.~W.}\ \bibnamefont {Taylor}}, \bibinfo {author}
  {\bibfnamefont {H.~J.}\ \bibnamefont {Kang}}, \bibinfo {author}
  {\bibfnamefont {R.}~\bibnamefont {Hu}},\ and\ \bibinfo {author}
  {\bibfnamefont {C.}~\bibnamefont {Petrovic}},\ }\bibfield  {title} {\bibinfo
  {title} {From incommensurate correlations to mesoscopic spin resonance {in
  YbRh}$_2${Si}$_2$},\ }\href {https://doi.org/10.1103/physrevlett.109.127201}
  {\bibfield  {journal} {\bibinfo  {journal} {Physical Review Letters}\
  }\textbf {\bibinfo {volume} {109}},\ \bibinfo {pages} {127201} (\bibinfo
  {year} {2012})}\BibitemShut {NoStop}%
\bibitem [{\citenamefont {Gegenwart}\ \emph {et~al.}(2005)\citenamefont
  {Gegenwart}, \citenamefont {Custers}, \citenamefont {Tokiwa}, \citenamefont
  {Geibel},\ and\ \citenamefont {Steglich}}]{Gegenwart2005}%
  \BibitemOpen
  \bibfield  {author} {\bibinfo {author} {\bibfnamefont {P.}~\bibnamefont
  {Gegenwart}}, \bibinfo {author} {\bibfnamefont {J.}~\bibnamefont {Custers}},
  \bibinfo {author} {\bibfnamefont {Y.}~\bibnamefont {Tokiwa}}, \bibinfo
  {author} {\bibfnamefont {C.}~\bibnamefont {Geibel}},\ and\ \bibinfo {author}
  {\bibfnamefont {F.}~\bibnamefont {Steglich}},\ }\bibfield  {title} {\bibinfo
  {title} {Ferromagnetic quantum critical fluctuations in
  {YbRh$_2$(Si$_{0.95}$Ge$_{0.05}$)$_2$}},\ }\href
  {https://doi.org/10.1103/physrevlett.94.076402} {\bibfield  {journal}
  {\bibinfo  {journal} {Physical Review Letters}\ }\textbf {\bibinfo {volume}
  {94}},\ \bibinfo {pages} {076402} (\bibinfo {year} {2005})}\BibitemShut
  {NoStop}%
\bibitem [{\citenamefont {Ishida}\ \emph {et~al.}(2002)\citenamefont {Ishida},
  \citenamefont {Okamoto}, \citenamefont {Kawasaki}, \citenamefont {Kitaoka},
  \citenamefont {Trovarelli}, \citenamefont {Geibel},\ and\ \citenamefont
  {Steglich}}]{Ishida2002}%
  \BibitemOpen
  \bibfield  {author} {\bibinfo {author} {\bibfnamefont {K.}~\bibnamefont
  {Ishida}}, \bibinfo {author} {\bibfnamefont {K.}~\bibnamefont {Okamoto}},
  \bibinfo {author} {\bibfnamefont {Y.}~\bibnamefont {Kawasaki}}, \bibinfo
  {author} {\bibfnamefont {Y.}~\bibnamefont {Kitaoka}}, \bibinfo {author}
  {\bibfnamefont {O.}~\bibnamefont {Trovarelli}}, \bibinfo {author}
  {\bibfnamefont {C.}~\bibnamefont {Geibel}},\ and\ \bibinfo {author}
  {\bibfnamefont {F.}~\bibnamefont {Steglich}},\ }\bibfield  {title} {\bibinfo
  {title} {{YbRh}$_2${Si}$_2$: Spin fluctuations in the vicinity of a quantum
  critical point at low magnetic field},\ }\href
  {https://doi.org/10.1103/physrevlett.89.107202} {\bibfield  {journal}
  {\bibinfo  {journal} {Physical Review Letters}\ }\textbf {\bibinfo {volume}
  {89}},\ \bibinfo {pages} {107202} (\bibinfo {year} {2002})}\BibitemShut
  {NoStop}%
\bibitem [{\citenamefont {Sichelschmidt}\ \emph {et~al.}(2003)\citenamefont
  {Sichelschmidt}, \citenamefont {Ivanshin}, \citenamefont {Ferstl},
  \citenamefont {Geibel},\ and\ \citenamefont {Steglich}}]{Sichelschmidt2003}%
  \BibitemOpen
  \bibfield  {author} {\bibinfo {author} {\bibfnamefont {J.}~\bibnamefont
  {Sichelschmidt}}, \bibinfo {author} {\bibfnamefont {V.~A.}\ \bibnamefont
  {Ivanshin}}, \bibinfo {author} {\bibfnamefont {J.}~\bibnamefont {Ferstl}},
  \bibinfo {author} {\bibfnamefont {C.}~\bibnamefont {Geibel}},\ and\ \bibinfo
  {author} {\bibfnamefont {F.}~\bibnamefont {Steglich}},\ }\bibfield  {title}
  {\bibinfo {title} {Low temperature electron spin resonance of the {Kondo} ion
  in a heavy fermion metal: {YbRh\textsubscript{2}Si\textsubscript{2}}},\
  }\href {https://doi.org/10.1103/physrevlett.91.156401} {\bibfield  {journal}
  {\bibinfo  {journal} {Physical Review Letters}\ }\textbf {\bibinfo {volume}
  {91}},\ \bibinfo {pages} {156401} (\bibinfo {year} {2003})}\BibitemShut
  {NoStop}%
\bibitem [{\citenamefont {Abrahams}\ and\ \citenamefont
  {W{\"{o}}lfle}(2008)}]{Abrahams2008}%
  \BibitemOpen
  \bibfield  {author} {\bibinfo {author} {\bibfnamefont {E.}~\bibnamefont
  {Abrahams}}\ and\ \bibinfo {author} {\bibfnamefont {P.}~\bibnamefont
  {W{\"{o}}lfle}},\ }\bibfield  {title} {\bibinfo {title} {Electron spin
  resonance in {Kondo} systems},\ }\href
  {https://doi.org/10.1103/physrevb.78.104423} {\bibfield  {journal} {\bibinfo
  {journal} {Physical Review B}\ }\textbf {\bibinfo {volume} {78}},\ \bibinfo
  {pages} {104423} (\bibinfo {year} {2008})}\BibitemShut {NoStop}%
\bibitem [{\citenamefont {W{\"{o}}lfle}\ and\ \citenamefont
  {Abrahams}(2009)}]{Woelfle2009}%
  \BibitemOpen
  \bibfield  {author} {\bibinfo {author} {\bibfnamefont {P.}~\bibnamefont
  {W{\"{o}}lfle}}\ and\ \bibinfo {author} {\bibfnamefont {E.}~\bibnamefont
  {Abrahams}},\ }\bibfield  {title} {\bibinfo {title} {Phenomenology of {ESR}
  in heavy-fermion systems: {Fermi-liquid and non-Fermi-liquid regimes}},\
  }\href {https://doi.org/10.1103/physrevb.80.235112} {\bibfield  {journal}
  {\bibinfo  {journal} {Physical Review B}\ }\textbf {\bibinfo {volume} {80}},\
  \bibinfo {pages} {235112} (\bibinfo {year} {2009})}\BibitemShut {NoStop}%
\bibitem [{\citenamefont {Si}\ \emph {et~al.}(2001)\citenamefont {Si},
  \citenamefont {Rabello}, \citenamefont {Ingersent},\ and\ \citenamefont
  {Smith}}]{Si2001}%
  \BibitemOpen
  \bibfield  {author} {\bibinfo {author} {\bibfnamefont {Q.}~\bibnamefont
  {Si}}, \bibinfo {author} {\bibfnamefont {S.}~\bibnamefont {Rabello}},
  \bibinfo {author} {\bibfnamefont {K.}~\bibnamefont {Ingersent}},\ and\
  \bibinfo {author} {\bibfnamefont {J.~L.}\ \bibnamefont {Smith}},\ }\bibfield
  {title} {\bibinfo {title} {Locally critical quantum phase transitions in
  strongly correlated metals},\ }\href {https://doi.org/10.1038/35101507}
  {\bibfield  {journal} {\bibinfo  {journal} {Nature}\ }\textbf {\bibinfo
  {volume} {413}},\ \bibinfo {pages} {804} (\bibinfo {year}
  {2001})}\BibitemShut {NoStop}%
\bibitem [{\citenamefont {Coleman}\ and\ \citenamefont
  {Schofield}(2005)}]{Coleman2005}%
  \BibitemOpen
  \bibfield  {author} {\bibinfo {author} {\bibfnamefont {P.}~\bibnamefont
  {Coleman}}\ and\ \bibinfo {author} {\bibfnamefont {A.~J.}\ \bibnamefont
  {Schofield}},\ }\bibfield  {title} {\bibinfo {title} {Quantum criticality},\
  }\href {https://doi.org/10.1038/nature03279} {\bibfield  {journal} {\bibinfo
  {journal} {Nature}\ }\textbf {\bibinfo {volume} {433}},\ \bibinfo {pages}
  {226} (\bibinfo {year} {2005})}\BibitemShut {NoStop}%
\bibitem [{\citenamefont {Si}\ and\ \citenamefont {Steglich}(2010)}]{Si2010}%
  \BibitemOpen
  \bibfield  {author} {\bibinfo {author} {\bibfnamefont {Q.}~\bibnamefont
  {Si}}\ and\ \bibinfo {author} {\bibfnamefont {F.}~\bibnamefont {Steglich}},\
  }\bibfield  {title} {\bibinfo {title} {Heavy {Fermions} and quantum phase
  transitions},\ }\href {https://doi.org/10.1126/science.1191195} {\bibfield
  {journal} {\bibinfo  {journal} {Science}\ }\textbf {\bibinfo {volume}
  {329}},\ \bibinfo {pages} {1161} (\bibinfo {year} {2010})}\BibitemShut
  {NoStop}%
\bibitem [{\citenamefont {Steglich}(2014)}]{Steglich2014}%
  \BibitemOpen
  \bibfield  {author} {\bibinfo {author} {\bibfnamefont {F.}~\bibnamefont
  {Steglich}},\ }\bibfield  {title} {\bibinfo {title} {Heavy fermions:
  superconductivity and its relationship to quantum criticality},\ }\href
  {https://doi.org/10.1080/14786435.2014.956835} {\bibfield  {journal}
  {\bibinfo  {journal} {Philosophical Magazine}\ }\textbf {\bibinfo {volume}
  {94}},\ \bibinfo {pages} {3259} (\bibinfo {year} {2014})}\BibitemShut
  {NoStop}%
\bibitem [{\citenamefont {Schubert}\ \emph {et~al.}(2019)\citenamefont
  {Schubert}, \citenamefont {Tokiwa}, \citenamefont {H{\"{u}}bner},
  \citenamefont {Mchalwat}, \citenamefont {Blumenr{\"{o}}ther}, \citenamefont
  {Jeevan},\ and\ \citenamefont {Gegenwart}}]{Schubert2019}%
  \BibitemOpen
  \bibfield  {author} {\bibinfo {author} {\bibfnamefont {M.~H.}\ \bibnamefont
  {Schubert}}, \bibinfo {author} {\bibfnamefont {Y.}~\bibnamefont {Tokiwa}},
  \bibinfo {author} {\bibfnamefont {S.~H.}\ \bibnamefont {H{\"{u}}bner}},
  \bibinfo {author} {\bibfnamefont {M.}~\bibnamefont {Mchalwat}}, \bibinfo
  {author} {\bibfnamefont {E.}~\bibnamefont {Blumenr{\"{o}}ther}}, \bibinfo
  {author} {\bibfnamefont {H.~S.}\ \bibnamefont {Jeevan}},\ and\ \bibinfo
  {author} {\bibfnamefont {P.}~\bibnamefont {Gegenwart}},\ }\bibfield  {title}
  {\bibinfo {title} {Tuning low-energy scales in {YbRh$_2$Si$_2$} by
  non-isoelectronic substitution and pressure},\ }\href
  {https://doi.org/10.1103/physrevresearch.1.032004} {\bibfield  {journal}
  {\bibinfo  {journal} {Physical Review Research}\ }\textbf {\bibinfo {volume}
  {1}},\ \bibinfo {pages} {032004(R)} (\bibinfo {year} {2019})}\BibitemShut
  {NoStop}%
\bibitem [{\citenamefont {Gegenwart}\ \emph {et~al.}(2007)\citenamefont
  {Gegenwart}, \citenamefont {Westerkamp}, \citenamefont {Krellner},
  \citenamefont {Tokiwa}, \citenamefont {Paschen}, \citenamefont {Geibel},
  \citenamefont {Steglich}, \citenamefont {Abrahams},\ and\ \citenamefont
  {Si}}]{Gegenwart2007}%
  \BibitemOpen
  \bibfield  {author} {\bibinfo {author} {\bibfnamefont {P.}~\bibnamefont
  {Gegenwart}}, \bibinfo {author} {\bibfnamefont {T.}~\bibnamefont
  {Westerkamp}}, \bibinfo {author} {\bibfnamefont {C.}~\bibnamefont
  {Krellner}}, \bibinfo {author} {\bibfnamefont {Y.}~\bibnamefont {Tokiwa}},
  \bibinfo {author} {\bibfnamefont {S.}~\bibnamefont {Paschen}}, \bibinfo
  {author} {\bibfnamefont {C.}~\bibnamefont {Geibel}}, \bibinfo {author}
  {\bibfnamefont {F.}~\bibnamefont {Steglich}}, \bibinfo {author}
  {\bibfnamefont {E.}~\bibnamefont {Abrahams}},\ and\ \bibinfo {author}
  {\bibfnamefont {Q.}~\bibnamefont {Si}},\ }\bibfield  {title} {\bibinfo
  {title} {Multiple energy scales at a quantum critical point},\ }\href
  {https://doi.org/10.1126/science.1136020} {\bibfield  {journal} {\bibinfo
  {journal} {Science}\ }\textbf {\bibinfo {volume} {315}},\ \bibinfo {pages}
  {969} (\bibinfo {year} {2007})}\BibitemShut {NoStop}%
\bibitem [{\citenamefont {Gegenwart}\ \emph {et~al.}(2008)\citenamefont
  {Gegenwart}, \citenamefont {Si},\ and\ \citenamefont
  {Steglich}}]{Gegenwart2008}%
  \BibitemOpen
  \bibfield  {author} {\bibinfo {author} {\bibfnamefont {P.}~\bibnamefont
  {Gegenwart}}, \bibinfo {author} {\bibfnamefont {Q.}~\bibnamefont {Si}},\ and\
  \bibinfo {author} {\bibfnamefont {F.}~\bibnamefont {Steglich}},\ }\bibfield
  {title} {\bibinfo {title} {Quantum criticality in heavy{\textendash}fermion
  metals},\ }\href {https://doi.org/10.1038/nphys892} {\bibfield  {journal}
  {\bibinfo  {journal} {Nature Physics}\ }\textbf {\bibinfo {volume} {4}},\
  \bibinfo {pages} {186} (\bibinfo {year} {2008})}\BibitemShut {NoStop}%
\bibitem [{\citenamefont {Paschen}\ and\ \citenamefont
  {Si}(2020)}]{Paschen2020}%
  \BibitemOpen
  \bibfield  {author} {\bibinfo {author} {\bibfnamefont {S.}~\bibnamefont
  {Paschen}}\ and\ \bibinfo {author} {\bibfnamefont {Q.}~\bibnamefont {Si}},\
  }\bibfield  {title} {\bibinfo {title} {Quantum phases driven by strong
  correlations},\ }\href {https://doi.org/10.1038/s42254-020-00262-6}
  {\bibfield  {journal} {\bibinfo  {journal} {Nature Reviews Physics}\ }\textbf
  {\bibinfo {volume} {3}},\ \bibinfo {pages} {9} (\bibinfo {year}
  {2020})}\BibitemShut {NoStop}%
\bibitem [{\citenamefont {Schuberth}\ \emph {et~al.}(2022)\citenamefont
  {Schuberth}, \citenamefont {Wirth},\ and\ \citenamefont
  {Steglich}}]{Schuberth2022}%
  \BibitemOpen
  \bibfield  {author} {\bibinfo {author} {\bibfnamefont {E.}~\bibnamefont
  {Schuberth}}, \bibinfo {author} {\bibfnamefont {S.}~\bibnamefont {Wirth}},\
  and\ \bibinfo {author} {\bibfnamefont {F.}~\bibnamefont {Steglich}},\
  }\bibfield  {title} {\bibinfo {title} {Nuclear-order-induced quantum
  criticality and {Heavy-Fermion} superconductivity at ultra-low temperatures
  in {YbRh}$_2${Si}$_2$},\ }\bibfield  {journal} {\bibinfo  {journal}
  {Frontiers in Electronic Materials}\ }\textbf {\bibinfo {volume} {2}},\ \href
  {https://doi.org/10.3389/femat.2022.869495} {10.3389/femat.2022.869495}
  (\bibinfo {year} {2022})\BibitemShut {NoStop}%
\bibitem [{\citenamefont {Abrahams}\ and\ \citenamefont
  {W{\"{o}}lfle}(2012)}]{Abrahams2012}%
  \BibitemOpen
  \bibfield  {author} {\bibinfo {author} {\bibfnamefont {E.}~\bibnamefont
  {Abrahams}}\ and\ \bibinfo {author} {\bibfnamefont {P.}~\bibnamefont
  {W{\"{o}}lfle}},\ }\bibfield  {title} {\bibinfo {title} {Critical
  quasiparticle theory applied to heavy fermion metals near an
  antiferromagnetic quantum phase transition},\ }\href
  {https://doi.org/10.1073/pnas.1200346109} {\bibfield  {journal} {\bibinfo
  {journal} {Proceedings of the National Academy of Sciences}\ }\textbf
  {\bibinfo {volume} {109}},\ \bibinfo {pages} {3238} (\bibinfo {year}
  {2012})}\BibitemShut {NoStop}%
\bibitem [{\citenamefont {Abrahams}\ \emph {et~al.}(2014)\citenamefont
  {Abrahams}, \citenamefont {Schmalian},\ and\ \citenamefont
  {W{\"{o}}lfle}}]{Abrahams2014}%
  \BibitemOpen
  \bibfield  {author} {\bibinfo {author} {\bibfnamefont {E.}~\bibnamefont
  {Abrahams}}, \bibinfo {author} {\bibfnamefont {J.}~\bibnamefont
  {Schmalian}},\ and\ \bibinfo {author} {\bibfnamefont {P.}~\bibnamefont
  {W{\"{o}}lfle}},\ }\bibfield  {title} {\bibinfo {title} {Strong-coupling
  theory of heavy-fermion criticality},\ }\href
  {https://doi.org/10.1103/physrevb.90.045105} {\bibfield  {journal} {\bibinfo
  {journal} {Physical Review B}\ }\textbf {\bibinfo {volume} {90}},\ \bibinfo
  {pages} {045105} (\bibinfo {year} {2014})}\BibitemShut {NoStop}%
\bibitem [{\citenamefont {W{\"{o}}lfle}\ and\ \citenamefont
  {Abrahams}(2015)}]{Woelfle2015}%
  \BibitemOpen
  \bibfield  {author} {\bibinfo {author} {\bibfnamefont {P.}~\bibnamefont
  {W{\"{o}}lfle}}\ and\ \bibinfo {author} {\bibfnamefont {E.}~\bibnamefont
  {Abrahams}},\ }\bibfield  {title} {\bibinfo {title} {Spin-flip scattering of
  critical quasiparticles and the phase diagram of
  {YbRh\textsubscript{2}Si\textsubscript{2}}},\ }\href
  {https://doi.org/10.1103/physrevb.92.155111} {\bibfield  {journal} {\bibinfo
  {journal} {Physical Review B}\ }\textbf {\bibinfo {volume} {92}},\ \bibinfo
  {pages} {155111} (\bibinfo {year} {2015})}\BibitemShut {NoStop}%
\bibitem [{\citenamefont {Custers}\ \emph {et~al.}(2003)\citenamefont
  {Custers}, \citenamefont {Gegenwart}, \citenamefont {Wilhelm}, \citenamefont
  {Neumaier}, \citenamefont {Tokiwa}, \citenamefont {Trovarelli}, \citenamefont
  {Geibel}, \citenamefont {Steglich}, \citenamefont {P{\'{e}}pin},\ and\
  \citenamefont {Coleman}}]{Custers2003a}%
  \BibitemOpen
  \bibfield  {author} {\bibinfo {author} {\bibfnamefont {J.}~\bibnamefont
  {Custers}}, \bibinfo {author} {\bibfnamefont {P.}~\bibnamefont {Gegenwart}},
  \bibinfo {author} {\bibfnamefont {H.}~\bibnamefont {Wilhelm}}, \bibinfo
  {author} {\bibfnamefont {K.}~\bibnamefont {Neumaier}}, \bibinfo {author}
  {\bibfnamefont {Y.}~\bibnamefont {Tokiwa}}, \bibinfo {author} {\bibfnamefont
  {O.}~\bibnamefont {Trovarelli}}, \bibinfo {author} {\bibfnamefont
  {C.}~\bibnamefont {Geibel}}, \bibinfo {author} {\bibfnamefont
  {F.}~\bibnamefont {Steglich}}, \bibinfo {author} {\bibfnamefont
  {C.}~\bibnamefont {P{\'{e}}pin}},\ and\ \bibinfo {author} {\bibfnamefont
  {P.}~\bibnamefont {Coleman}},\ }\bibfield  {title} {\bibinfo {title} {The
  break-up of heavy electrons at a quantum critical point},\ }\href
  {https://doi.org/10.1038/nature01774} {\bibfield  {journal} {\bibinfo
  {journal} {Nature}\ }\textbf {\bibinfo {volume} {424}},\ \bibinfo {pages}
  {524} (\bibinfo {year} {2003})}\BibitemShut {NoStop}%
\bibitem [{\citenamefont {Lausberg}\ \emph {et~al.}(2013)\citenamefont
  {Lausberg}, \citenamefont {Hannaske}, \citenamefont {Steppke}, \citenamefont
  {Steinke}, \citenamefont {Gruner}, \citenamefont {Pedrero}, \citenamefont
  {Krellner}, \citenamefont {Klingner}, \citenamefont {Brando}, \citenamefont
  {Geibel},\ and\ \citenamefont {Steglich}}]{Lausberg2013}%
  \BibitemOpen
  \bibfield  {author} {\bibinfo {author} {\bibfnamefont {S.}~\bibnamefont
  {Lausberg}}, \bibinfo {author} {\bibfnamefont {A.}~\bibnamefont {Hannaske}},
  \bibinfo {author} {\bibfnamefont {A.}~\bibnamefont {Steppke}}, \bibinfo
  {author} {\bibfnamefont {L.}~\bibnamefont {Steinke}}, \bibinfo {author}
  {\bibfnamefont {T.}~\bibnamefont {Gruner}}, \bibinfo {author} {\bibfnamefont
  {L.}~\bibnamefont {Pedrero}}, \bibinfo {author} {\bibfnamefont
  {C.}~\bibnamefont {Krellner}}, \bibinfo {author} {\bibfnamefont
  {C.}~\bibnamefont {Klingner}}, \bibinfo {author} {\bibfnamefont
  {M.}~\bibnamefont {Brando}}, \bibinfo {author} {\bibfnamefont
  {C.}~\bibnamefont {Geibel}},\ and\ \bibinfo {author} {\bibfnamefont
  {F.}~\bibnamefont {Steglich}},\ }\bibfield  {title} {\bibinfo {title} {Doped
  {YbRh\textsubscript{2}Si\textsubscript{2}: Not} only ferromagnetic
  correlations but ferromagnetic order},\ }\href
  {https://doi.org/10.1103/physrevlett.110.256402} {\bibfield  {journal}
  {\bibinfo  {journal} {Physical Review Letters}\ }\textbf {\bibinfo {volume}
  {110}},\ \bibinfo {pages} {256402} (\bibinfo {year} {2013})}\BibitemShut
  {NoStop}%
\bibitem [{\citenamefont {Nguyen}\ \emph {et~al.}(2021)\citenamefont {Nguyen},
  \citenamefont {Sidorenko}, \citenamefont {Taupin}, \citenamefont {Knebel},
  \citenamefont {Lapertot}, \citenamefont {Schuberth},\ and\ \citenamefont
  {Paschen}}]{Nguyen2021}%
  \BibitemOpen
  \bibfield  {author} {\bibinfo {author} {\bibfnamefont {D.}~\bibnamefont
  {Nguyen}}, \bibinfo {author} {\bibfnamefont {A.}~\bibnamefont {Sidorenko}},
  \bibinfo {author} {\bibfnamefont {M.}~\bibnamefont {Taupin}}, \bibinfo
  {author} {\bibfnamefont {G.}~\bibnamefont {Knebel}}, \bibinfo {author}
  {\bibfnamefont {G.}~\bibnamefont {Lapertot}}, \bibinfo {author}
  {\bibfnamefont {E.}~\bibnamefont {Schuberth}},\ and\ \bibinfo {author}
  {\bibfnamefont {S.}~\bibnamefont {Paschen}},\ }\bibfield  {title} {\bibinfo
  {title} {Superconductivity in an extreme strange metal},\ }\bibfield
  {journal} {\bibinfo  {journal} {Nature Communications}\ }\textbf {\bibinfo
  {volume} {12}},\ \href {https://doi.org/10.1038/s41467-021-24670-z}
  {10.1038/s41467-021-24670-z} (\bibinfo {year} {2021})\BibitemShut {NoStop}%
\bibitem [{\citenamefont {Eisenlohr}\ and\ \citenamefont
  {Vojta}(2021)}]{Eisenlohr2021}%
  \BibitemOpen
  \bibfield  {author} {\bibinfo {author} {\bibfnamefont {H.}~\bibnamefont
  {Eisenlohr}}\ and\ \bibinfo {author} {\bibfnamefont {M.}~\bibnamefont
  {Vojta}},\ }\bibfield  {title} {\bibinfo {title} {Limits to magnetic quantum
  criticality from nuclear spins},\ }\href
  {https://doi.org/10.1103/physrevb.103.064405} {\bibfield  {journal} {\bibinfo
   {journal} {Physical Review B}\ }\textbf {\bibinfo {volume} {103}},\ \bibinfo
  {pages} {064405} (\bibinfo {year} {2021})}\BibitemShut {NoStop}%
\bibitem [{\citenamefont {Libersky}\ \emph {et~al.}(2021)\citenamefont
  {Libersky}, \citenamefont {McKenzie}, \citenamefont {Silevitch},
  \citenamefont {Stamp},\ and\ \citenamefont {Rosenbaum}}]{Libersky2021}%
  \BibitemOpen
  \bibfield  {author} {\bibinfo {author} {\bibfnamefont {M.}~\bibnamefont
  {Libersky}}, \bibinfo {author} {\bibfnamefont {R.~D.}\ \bibnamefont
  {McKenzie}}, \bibinfo {author} {\bibfnamefont {D.~M.}\ \bibnamefont
  {Silevitch}}, \bibinfo {author} {\bibfnamefont {P.~C.~E.}\ \bibnamefont
  {Stamp}},\ and\ \bibinfo {author} {\bibfnamefont {T.~F.}\ \bibnamefont
  {Rosenbaum}},\ }\bibfield  {title} {\bibinfo {title} {Direct observation of
  collective electronuclear modes about a quantum critical point},\ }\href
  {https://doi.org/10.1103/physrevlett.127.207202} {\bibfield  {journal}
  {\bibinfo  {journal} {Physical Review Letters}\ }\textbf {\bibinfo {volume}
  {127}},\ \bibinfo {pages} {207202} (\bibinfo {year} {2021})}\BibitemShut
  {NoStop}%
\bibitem [{\citenamefont {Brando}\ \emph {et~al.}(2013)\citenamefont {Brando},
  \citenamefont {Pedrero}, \citenamefont {Westerkamp}, \citenamefont
  {Krellner}, \citenamefont {Gegenwart}, \citenamefont {Geibel},\ and\
  \citenamefont {Steglich}}]{Brando2013}%
  \BibitemOpen
  \bibfield  {author} {\bibinfo {author} {\bibfnamefont {M.}~\bibnamefont
  {Brando}}, \bibinfo {author} {\bibfnamefont {L.}~\bibnamefont {Pedrero}},
  \bibinfo {author} {\bibfnamefont {T.}~\bibnamefont {Westerkamp}}, \bibinfo
  {author} {\bibfnamefont {C.}~\bibnamefont {Krellner}}, \bibinfo {author}
  {\bibfnamefont {P.}~\bibnamefont {Gegenwart}}, \bibinfo {author}
  {\bibfnamefont {C.}~\bibnamefont {Geibel}},\ and\ \bibinfo {author}
  {\bibfnamefont {F.}~\bibnamefont {Steglich}},\ }\bibfield  {title} {\bibinfo
  {title} {Magnetization study of the energy scales in {YbRh}$_2${Si}$_2$ under
  chemical pressure},\ }\href {https://doi.org/10.1002/pssb.201200771}
  {\bibfield  {journal} {\bibinfo  {journal} {Physica Status Solidi (B)}\
  }\textbf {\bibinfo {volume} {250}},\ \bibinfo {pages} {485} (\bibinfo {year}
  {2013})}\BibitemShut {NoStop}%
\bibitem [{\citenamefont {Casey}\ \emph {et~al.}(2014)\citenamefont {Casey},
  \citenamefont {Arnold}, \citenamefont {Levitin}, \citenamefont {Lusher},
  \citenamefont {Ny{\'{e}}ki}, \citenamefont {Saunders}, \citenamefont
  {Shibahara}, \citenamefont {van~der Vliet}, \citenamefont {Yager},
  \citenamefont {Drung}, \citenamefont {Schurig}, \citenamefont {Batey},
  \citenamefont {Cuthbert},\ and\ \citenamefont {Matthews}}]{Casey2014}%
  \BibitemOpen
  \bibfield  {author} {\bibinfo {author} {\bibfnamefont {A.}~\bibnamefont
  {Casey}}, \bibinfo {author} {\bibfnamefont {F.}~\bibnamefont {Arnold}},
  \bibinfo {author} {\bibfnamefont {L.~V.}\ \bibnamefont {Levitin}}, \bibinfo
  {author} {\bibfnamefont {C.~P.}\ \bibnamefont {Lusher}}, \bibinfo {author}
  {\bibfnamefont {J.}~\bibnamefont {Ny{\'{e}}ki}}, \bibinfo {author}
  {\bibfnamefont {J.}~\bibnamefont {Saunders}}, \bibinfo {author}
  {\bibfnamefont {A.}~\bibnamefont {Shibahara}}, \bibinfo {author}
  {\bibfnamefont {H.}~\bibnamefont {van~der Vliet}}, \bibinfo {author}
  {\bibfnamefont {B.}~\bibnamefont {Yager}}, \bibinfo {author} {\bibfnamefont
  {D.}~\bibnamefont {Drung}}, \bibinfo {author} {\bibfnamefont
  {T.}~\bibnamefont {Schurig}}, \bibinfo {author} {\bibfnamefont
  {G.}~\bibnamefont {Batey}}, \bibinfo {author} {\bibfnamefont {M.~N.}\
  \bibnamefont {Cuthbert}},\ and\ \bibinfo {author} {\bibfnamefont {A.~J.}\
  \bibnamefont {Matthews}},\ }\bibfield  {title} {\bibinfo {title} {Current
  sensing noise thermometry: a fast practical solution to low temperature
  measurement},\ }\href {https://doi.org/10.1007/s10909-014-1147-z} {\bibfield
  {journal} {\bibinfo  {journal} {Journal of Low Temperature Physics}\ }\textbf
  {\bibinfo {volume} {175}},\ \bibinfo {pages} {764–775} (\bibinfo {year}
  {2014})}\BibitemShut {NoStop}%
\bibitem [{\citenamefont {Levitin}\ \emph {et~al.}(2022)\citenamefont
  {Levitin}, \citenamefont {van~der Vliet}, \citenamefont {Theisen},
  \citenamefont {Dimitriadis}, \citenamefont {Lucas}, \citenamefont {Corcoles},
  \citenamefont {Ny{\'{e}}ki}, \citenamefont {Casey}, \citenamefont {Creeth},
  \citenamefont {Farrer}, \citenamefont {Ritchie}, \citenamefont {Nicholls},\
  and\ \citenamefont {Saunders}}]{Levitin2022a}%
  \BibitemOpen
  \bibfield  {author} {\bibinfo {author} {\bibfnamefont {L.}~\bibnamefont
  {Levitin}}, \bibinfo {author} {\bibfnamefont {H.}~\bibnamefont {van~der
  Vliet}}, \bibinfo {author} {\bibfnamefont {T.}~\bibnamefont {Theisen}},
  \bibinfo {author} {\bibfnamefont {S.}~\bibnamefont {Dimitriadis}}, \bibinfo
  {author} {\bibfnamefont {M.}~\bibnamefont {Lucas}}, \bibinfo {author}
  {\bibfnamefont {A.}~\bibnamefont {Corcoles}}, \bibinfo {author}
  {\bibfnamefont {J.}~\bibnamefont {Ny{\'{e}}ki}}, \bibinfo {author}
  {\bibfnamefont {A.}~\bibnamefont {Casey}}, \bibinfo {author} {\bibfnamefont
  {G.}~\bibnamefont {Creeth}}, \bibinfo {author} {\bibfnamefont
  {I.}~\bibnamefont {Farrer}}, \bibinfo {author} {\bibfnamefont
  {D.}~\bibnamefont {Ritchie}}, \bibinfo {author} {\bibfnamefont
  {J.}~\bibnamefont {Nicholls}},\ and\ \bibinfo {author} {\bibfnamefont
  {J.}~\bibnamefont {Saunders}},\ }\bibfield  {title} {\bibinfo {title}
  {Cooling low-dimensional electron systems into the microkelvin regime},\
  }\href {https://doi.org/10.1038/s41467-022-28222-x} {\bibfield  {journal}
  {\bibinfo  {journal} {Nature Communications}\ }\textbf {\bibinfo {volume}
  {13}},\ \bibinfo {pages} {1} (\bibinfo {year} {2022})}\BibitemShut {NoStop}%
\bibitem [{\citenamefont {Krellner}\ \emph {et~al.}(2012)\citenamefont
  {Krellner}, \citenamefont {Taube}, \citenamefont {Westerkamp}, \citenamefont
  {Hossain},\ and\ \citenamefont {Geibel}}]{Krellner2012}%
  \BibitemOpen
  \bibfield  {author} {\bibinfo {author} {\bibfnamefont {C.}~\bibnamefont
  {Krellner}}, \bibinfo {author} {\bibfnamefont {S.}~\bibnamefont {Taube}},
  \bibinfo {author} {\bibfnamefont {T.}~\bibnamefont {Westerkamp}}, \bibinfo
  {author} {\bibfnamefont {Z.}~\bibnamefont {Hossain}},\ and\ \bibinfo {author}
  {\bibfnamefont {C.}~\bibnamefont {Geibel}},\ }\bibfield  {title} {\bibinfo
  {title} {Single-crystal growth of {YbRh$_2$Si$_2$} and {YbIr$_2$Si$_2$}},\
  }\href {https://doi.org/10.1080/14786435.2012.669066} {\bibfield  {journal}
  {\bibinfo  {journal} {Philosophical Magazine}\ }\textbf {\bibinfo {volume}
  {92}},\ \bibinfo {pages} {2508} (\bibinfo {year} {2012})}\BibitemShut
  {NoStop}%
\bibitem [{\citenamefont {Knebel}\ \emph {et~al.}(2006)\citenamefont {Knebel},
  \citenamefont {Boursier}, \citenamefont {Hassinger}, \citenamefont
  {Lapertot}, \citenamefont {Niklowitz}, \citenamefont {Pourret}, \citenamefont
  {Salce}, \citenamefont {Sanchez}, \citenamefont {Sheikin}, \citenamefont
  {Bonville}, \citenamefont {Harima},\ and\ \citenamefont
  {Flouquet}}]{Knebel2006}%
  \BibitemOpen
  \bibfield  {author} {\bibinfo {author} {\bibfnamefont {G.}~\bibnamefont
  {Knebel}}, \bibinfo {author} {\bibfnamefont {R.}~\bibnamefont {Boursier}},
  \bibinfo {author} {\bibfnamefont {E.}~\bibnamefont {Hassinger}}, \bibinfo
  {author} {\bibfnamefont {G.}~\bibnamefont {Lapertot}}, \bibinfo {author}
  {\bibfnamefont {P.~G.}\ \bibnamefont {Niklowitz}}, \bibinfo {author}
  {\bibfnamefont {A.}~\bibnamefont {Pourret}}, \bibinfo {author} {\bibfnamefont
  {B.}~\bibnamefont {Salce}}, \bibinfo {author} {\bibfnamefont {J.~P.}\
  \bibnamefont {Sanchez}}, \bibinfo {author} {\bibfnamefont {I.}~\bibnamefont
  {Sheikin}}, \bibinfo {author} {\bibfnamefont {P.}~\bibnamefont {Bonville}},
  \bibinfo {author} {\bibfnamefont {H.}~\bibnamefont {Harima}},\ and\ \bibinfo
  {author} {\bibfnamefont {J.}~\bibnamefont {Flouquet}},\ }\bibfield  {title}
  {\bibinfo {title} {Localization of 4f state in {YbRh}$_2${Si}$_2$ under
  magnetic field and high pressure: Comparison with {CeRh}$_2${Si}$_2$},\
  }\href {https://doi.org/10.1143/jpsj.75.114709} {\bibfield  {journal}
  {\bibinfo  {journal} {Journal of the Physical Society of Japan}\ }\textbf
  {\bibinfo {volume} {75}},\ \bibinfo {pages} {114709} (\bibinfo {year}
  {2006})}\BibitemShut {NoStop}%
\bibitem [{\citenamefont {Kondo}(1961)}]{Kondo1961}%
  \BibitemOpen
  \bibfield  {author} {\bibinfo {author} {\bibfnamefont {J.}~\bibnamefont
  {Kondo}},\ }\bibfield  {title} {\bibinfo {title} {Internal magnetic field in
  rare earth metals},\ }\href {https://doi.org/10.1143/jpsj.16.1690} {\bibfield
   {journal} {\bibinfo  {journal} {Journal of the Physical Society of Japan}\
  }\textbf {\bibinfo {volume} {16}},\ \bibinfo {pages} {1690} (\bibinfo {year}
  {1961})}\BibitemShut {NoStop}%
\bibitem [{\citenamefont {Bonville}\ \emph {et~al.}(1984)\citenamefont
  {Bonville}, \citenamefont {Imbert}, \citenamefont {J{\'{e}}hanno},
  \citenamefont {Gonzalez{\textendash}Jimenez},\ and\ \citenamefont
  {Hartmann{\textendash}Boutron}}]{Bonville1984}%
  \BibitemOpen
  \bibfield  {author} {\bibinfo {author} {\bibfnamefont {P.}~\bibnamefont
  {Bonville}}, \bibinfo {author} {\bibfnamefont {P.}~\bibnamefont {Imbert}},
  \bibinfo {author} {\bibfnamefont {G.}~\bibnamefont {J{\'{e}}hanno}}, \bibinfo
  {author} {\bibfnamefont {F.}~\bibnamefont {Gonzalez{\textendash}Jimenez}},\
  and\ \bibinfo {author} {\bibfnamefont {F.}~\bibnamefont
  {Hartmann{\textendash}Boutron}},\ }\bibfield  {title} {\bibinfo {title}
  {Emission {M{\"{o}}ssbauer} spectroscopy and relaxation measurements in
  hyperfine levels out of thermal equilibrium:
  Very{\textendash}low{\textendash}temperature experiments on the {Kondo}
  {alloy Au$^{170}$Yb}},\ }\href {https://doi.org/10.1103/physrevb.30.3672}
  {\bibfield  {journal} {\bibinfo  {journal} {Physical Review B}\ }\textbf
  {\bibinfo {volume} {30}},\ \bibinfo {pages} {3672} (\bibinfo {year}
  {1984})}\BibitemShut {NoStop}%
\bibitem [{\citenamefont {Bonville}\ \emph {et~al.}(1991)\citenamefont
  {Bonville}, \citenamefont {Hodges}, \citenamefont {Imbert}, \citenamefont
  {J{\'{e}}hanno}, \citenamefont {Jaccard},\ and\ \citenamefont
  {Sierro}}]{Bonville1991}%
  \BibitemOpen
  \bibfield  {author} {\bibinfo {author} {\bibfnamefont {P.}~\bibnamefont
  {Bonville}}, \bibinfo {author} {\bibfnamefont {J.~A.}\ \bibnamefont
  {Hodges}}, \bibinfo {author} {\bibfnamefont {P.}~\bibnamefont {Imbert}},
  \bibinfo {author} {\bibfnamefont {G.}~\bibnamefont {J{\'{e}}hanno}}, \bibinfo
  {author} {\bibfnamefont {D.}~\bibnamefont {Jaccard}},\ and\ \bibinfo {author}
  {\bibfnamefont {J.}~\bibnamefont {Sierro}},\ }\bibfield  {title} {\bibinfo
  {title} {Magnetic ordering and paramagnetic relaxation of {Yb}$^{3+}$ in
  {YbNi}$_2${Si}$_2$},\ }\href {https://doi.org/10.1016/0304-8853(91)90178-d}
  {\bibfield  {journal} {\bibinfo  {journal} {Journal of Magnetism and Magnetic
  Materials}\ }\textbf {\bibinfo {volume} {97}},\ \bibinfo {pages} {178}
  (\bibinfo {year} {1991})}\BibitemShut {NoStop}%
\bibitem [{\citenamefont {Nowik}\ and\ \citenamefont {Ofer}(1968)}]{Nowik1968}%
  \BibitemOpen
  \bibfield  {author} {\bibinfo {author} {\bibfnamefont {I.}~\bibnamefont
  {Nowik}}\ and\ \bibinfo {author} {\bibfnamefont {S.}~\bibnamefont {Ofer}},\
  }\bibfield  {title} {\bibinfo {title} {{Mössbauer} studies of $^{170}${Yb}
  in several paramagnetic salts},\ }\href
  {https://doi.org/10.1016/0022-3697(68)90007-3} {\bibfield  {journal}
  {\bibinfo  {journal} {Journal of Physics and Chemistry of Solids}\ }\textbf
  {\bibinfo {volume} {29}},\ \bibinfo {pages} {2117} (\bibinfo {year}
  {1968})}\BibitemShut {NoStop}%
\bibitem [{\citenamefont {Plessel}\ \emph {et~al.}(2003)\citenamefont
  {Plessel}, \citenamefont {Abd{\textendash}Elmeguid}, \citenamefont {Sanchez},
  \citenamefont {Knebel}, \citenamefont {Geibel}, \citenamefont {Trovarelli},\
  and\ \citenamefont {Steglich}}]{Plessel2003}%
  \BibitemOpen
  \bibfield  {author} {\bibinfo {author} {\bibfnamefont {J.}~\bibnamefont
  {Plessel}}, \bibinfo {author} {\bibfnamefont {M.~M.}\ \bibnamefont
  {Abd{\textendash}Elmeguid}}, \bibinfo {author} {\bibfnamefont {J.~P.}\
  \bibnamefont {Sanchez}}, \bibinfo {author} {\bibfnamefont {G.}~\bibnamefont
  {Knebel}}, \bibinfo {author} {\bibfnamefont {C.}~\bibnamefont {Geibel}},
  \bibinfo {author} {\bibfnamefont {O.}~\bibnamefont {Trovarelli}},\ and\
  \bibinfo {author} {\bibfnamefont {F.}~\bibnamefont {Steglich}},\ }\bibfield
  {title} {\bibinfo {title} {Unusual behavior of the low{\textendash}moment
  magnetic ground state of {YbRh}$_2${Si}$_2$ under high pressure},\ }\href
  {https://doi.org/10.1103/physrevb.67.180403} {\bibfield  {journal} {\bibinfo
  {journal} {Physical Review B}\ }\textbf {\bibinfo {volume} {67}},\ \bibinfo
  {pages} {180403(R)} (\bibinfo {year} {2003})}\BibitemShut {NoStop}%
\bibitem [{\citenamefont {Flouquet}\ and\ \citenamefont
  {Brewer}(1975)}]{Flouquet1975}%
  \BibitemOpen
  \bibfield  {author} {\bibinfo {author} {\bibfnamefont {J.}~\bibnamefont
  {Flouquet}}\ and\ \bibinfo {author} {\bibfnamefont {W.~D.}\ \bibnamefont
  {Brewer}},\ }\bibfield  {title} {\bibinfo {title} {Hyperfine interaction
  studies of local moments in metals},\ }\href
  {https://doi.org/10.1088/0031-8949/11/3-4/013} {\bibfield  {journal}
  {\bibinfo  {journal} {Physica Scripta}\ }\textbf {\bibinfo {volume} {11}},\
  \bibinfo {pages} {199} (\bibinfo {year} {1975})}\BibitemShut {NoStop}%
\bibitem [{\citenamefont {Flouquet}(1978)}]{Flouquet1978}%
  \BibitemOpen
  \bibfield  {author} {\bibinfo {author} {\bibfnamefont {J.}~\bibnamefont
  {Flouquet}},\ }\bibfield  {title} {\bibinfo {title} {Kondo coupling,
  hyperfine and exchange interactions},\ }\href
  {https://doi.org/10.1051/jphyscol:19786592} {\bibfield  {journal} {\bibinfo
  {journal} {Le Journal de Physique Colloques}\ }\textbf {\bibinfo {volume}
  {39}},\ \bibinfo {pages} {C6} (\bibinfo {year} {1978})}\BibitemShut {NoStop}%
\bibitem [{\citenamefont {Steppke}\ \emph {et~al.}(2010)\citenamefont
  {Steppke}, \citenamefont {Brando}, \citenamefont {Oeschler}, \citenamefont
  {Krellner}, \citenamefont {Geibel},\ and\ \citenamefont
  {Steglich}}]{Steppke2010}%
  \BibitemOpen
  \bibfield  {author} {\bibinfo {author} {\bibfnamefont {A.}~\bibnamefont
  {Steppke}}, \bibinfo {author} {\bibfnamefont {M.}~\bibnamefont {Brando}},
  \bibinfo {author} {\bibfnamefont {N.}~\bibnamefont {Oeschler}}, \bibinfo
  {author} {\bibfnamefont {C.}~\bibnamefont {Krellner}}, \bibinfo {author}
  {\bibfnamefont {C.}~\bibnamefont {Geibel}},\ and\ \bibinfo {author}
  {\bibfnamefont {F.}~\bibnamefont {Steglich}},\ }\bibfield  {title} {\bibinfo
  {title} {Nuclear contribution to the specific heat of
  {Yb(Rh}$_{0.93}${Co}$_{0.07}$)$_2${Si}$_2$},\ }\href
  {https://doi.org/10.1002/pssb.200983062} {\bibfield  {journal} {\bibinfo
  {journal} {Physica Status Solidi (B)}\ }\textbf {\bibinfo {volume} {247}},\
  \bibinfo {pages} {737} (\bibinfo {year} {2010})}\BibitemShut {NoStop}%
\bibitem [{\citenamefont {Krellner}\ \emph {et~al.}(2009)\citenamefont
  {Krellner}, \citenamefont {Hartmann}, \citenamefont {Pikul}, \citenamefont
  {Oeschler}, \citenamefont {Donath}, \citenamefont {Geibel}, \citenamefont
  {Steglich},\ and\ \citenamefont {Wosnitza}}]{Krellner2009}%
  \BibitemOpen
  \bibfield  {author} {\bibinfo {author} {\bibfnamefont {C.}~\bibnamefont
  {Krellner}}, \bibinfo {author} {\bibfnamefont {S.}~\bibnamefont {Hartmann}},
  \bibinfo {author} {\bibfnamefont {A.}~\bibnamefont {Pikul}}, \bibinfo
  {author} {\bibfnamefont {N.}~\bibnamefont {Oeschler}}, \bibinfo {author}
  {\bibfnamefont {J.~G.}\ \bibnamefont {Donath}}, \bibinfo {author}
  {\bibfnamefont {C.}~\bibnamefont {Geibel}}, \bibinfo {author} {\bibfnamefont
  {F.}~\bibnamefont {Steglich}},\ and\ \bibinfo {author} {\bibfnamefont
  {J.}~\bibnamefont {Wosnitza}},\ }\bibfield  {title} {\bibinfo {title}
  {Violation of critical universality at the antiferromagnetic phase transition
  of {YbRh}$_2${Si}$_2$},\ }\href
  {https://doi.org/10.1103/physrevlett.102.196402} {\bibfield  {journal}
  {\bibinfo  {journal} {Physical Review Letters}\ }\textbf {\bibinfo {volume}
  {102}},\ \bibinfo {pages} {196402} (\bibinfo {year} {2009})}\BibitemShut
  {NoStop}%
\bibitem [{\citenamefont {Friedemann}\ \emph {et~al.}(2009)\citenamefont
  {Friedemann}, \citenamefont {Westerkamp}, \citenamefont {Brando},
  \citenamefont {Oeschler}, \citenamefont {Wirth}, \citenamefont {Gegenwart},
  \citenamefont {Krellner}, \citenamefont {Geibel},\ and\ \citenamefont
  {Steglich}}]{Friedemann2009}%
  \BibitemOpen
  \bibfield  {author} {\bibinfo {author} {\bibfnamefont {S.}~\bibnamefont
  {Friedemann}}, \bibinfo {author} {\bibfnamefont {T.}~\bibnamefont
  {Westerkamp}}, \bibinfo {author} {\bibfnamefont {M.}~\bibnamefont {Brando}},
  \bibinfo {author} {\bibfnamefont {N.}~\bibnamefont {Oeschler}}, \bibinfo
  {author} {\bibfnamefont {S.}~\bibnamefont {Wirth}}, \bibinfo {author}
  {\bibfnamefont {P.}~\bibnamefont {Gegenwart}}, \bibinfo {author}
  {\bibfnamefont {C.}~\bibnamefont {Krellner}}, \bibinfo {author}
  {\bibfnamefont {C.}~\bibnamefont {Geibel}},\ and\ \bibinfo {author}
  {\bibfnamefont {F.}~\bibnamefont {Steglich}},\ }\bibfield  {title} {\bibinfo
  {title} {Detaching the antiferromagnetic quantum critical point from the
  {Fermi}-surface reconstruction in
  {YbRh\textsubscript{2}Si\textsubscript{2}}},\ }\href
  {https://doi.org/10.1038/nphys1299} {\bibfield  {journal} {\bibinfo
  {journal} {Nature Physics}\ }\textbf {\bibinfo {volume} {5}},\ \bibinfo
  {pages} {465} (\bibinfo {year} {2009})}\BibitemShut {NoStop}%
\bibitem [{\citenamefont {Steinke}\ \emph {et~al.}(2017)\citenamefont
  {Steinke}, \citenamefont {Schuberth}, \citenamefont {Lausberg}, \citenamefont
  {Tippmann}, \citenamefont {Steppke}, \citenamefont {Krellner}, \citenamefont
  {Geibel}, \citenamefont {Steglich},\ and\ \citenamefont
  {Brando}}]{Steinke2017}%
  \BibitemOpen
  \bibfield  {author} {\bibinfo {author} {\bibfnamefont {L.}~\bibnamefont
  {Steinke}}, \bibinfo {author} {\bibfnamefont {E.}~\bibnamefont {Schuberth}},
  \bibinfo {author} {\bibfnamefont {S.}~\bibnamefont {Lausberg}}, \bibinfo
  {author} {\bibfnamefont {M.}~\bibnamefont {Tippmann}}, \bibinfo {author}
  {\bibfnamefont {A.}~\bibnamefont {Steppke}}, \bibinfo {author} {\bibfnamefont
  {C.}~\bibnamefont {Krellner}}, \bibinfo {author} {\bibfnamefont
  {C.}~\bibnamefont {Geibel}}, \bibinfo {author} {\bibfnamefont
  {F.}~\bibnamefont {Steglich}},\ and\ \bibinfo {author} {\bibfnamefont
  {M.}~\bibnamefont {Brando}},\ }\bibfield  {title} {\bibinfo {title}
  {Ultra{\textendash}low temperature ac susceptibility of the
  heavy{\textendash}fermion superconductor {YbRh$_2$Si$_2$}},\ }\href
  {https://doi.org/10.1088/1742-6596/807/5/052007} {\bibfield  {journal}
  {\bibinfo  {journal} {Journal of Physics: Conference Series}\ }\textbf
  {\bibinfo {volume} {807}},\ \bibinfo {pages} {052007} (\bibinfo {year}
  {2017})}\BibitemShut {NoStop}%
\bibitem [{\citenamefont {Smidman}\ \emph {et~al.}(2018)\citenamefont
  {Smidman}, \citenamefont {Stockert}, \citenamefont {Arndt}, \citenamefont
  {Pang}, \citenamefont {Jiao}, \citenamefont {Yuan}, \citenamefont {Vieyra},
  \citenamefont {Kitagawa}, \citenamefont {Ishida}, \citenamefont {Fujiwara},
  \citenamefont {Kobayashi}, \citenamefont {Schuberth}, \citenamefont
  {Tippmann}, \citenamefont {Steinke}, \citenamefont {Lausberg}, \citenamefont
  {Steppke}, \citenamefont {Brando}, \citenamefont {Pfau}, \citenamefont
  {Stockert}, \citenamefont {Sun}, \citenamefont {Friedemann}, \citenamefont
  {Wirth}, \citenamefont {Krellner}, \citenamefont {Kirchner}, \citenamefont
  {Nica}, \citenamefont {Yu}, \citenamefont {Si},\ and\ \citenamefont
  {Steglich}}]{Smidman2018}%
  \BibitemOpen
  \bibfield  {author} {\bibinfo {author} {\bibfnamefont {M.}~\bibnamefont
  {Smidman}}, \bibinfo {author} {\bibfnamefont {O.}~\bibnamefont {Stockert}},
  \bibinfo {author} {\bibfnamefont {J.}~\bibnamefont {Arndt}}, \bibinfo
  {author} {\bibfnamefont {G.~M.}\ \bibnamefont {Pang}}, \bibinfo {author}
  {\bibfnamefont {L.}~\bibnamefont {Jiao}}, \bibinfo {author} {\bibfnamefont
  {H.~Q.}\ \bibnamefont {Yuan}}, \bibinfo {author} {\bibfnamefont {H.~A.}\
  \bibnamefont {Vieyra}}, \bibinfo {author} {\bibfnamefont {S.}~\bibnamefont
  {Kitagawa}}, \bibinfo {author} {\bibfnamefont {K.}~\bibnamefont {Ishida}},
  \bibinfo {author} {\bibfnamefont {K.}~\bibnamefont {Fujiwara}}, \bibinfo
  {author} {\bibfnamefont {T.~C.}\ \bibnamefont {Kobayashi}}, \bibinfo {author}
  {\bibfnamefont {E.}~\bibnamefont {Schuberth}}, \bibinfo {author}
  {\bibfnamefont {M.}~\bibnamefont {Tippmann}}, \bibinfo {author}
  {\bibfnamefont {L.}~\bibnamefont {Steinke}}, \bibinfo {author} {\bibfnamefont
  {S.}~\bibnamefont {Lausberg}}, \bibinfo {author} {\bibfnamefont
  {A.}~\bibnamefont {Steppke}}, \bibinfo {author} {\bibfnamefont
  {M.}~\bibnamefont {Brando}}, \bibinfo {author} {\bibfnamefont
  {H.}~\bibnamefont {Pfau}}, \bibinfo {author} {\bibfnamefont {U.}~\bibnamefont
  {Stockert}}, \bibinfo {author} {\bibfnamefont {P.}~\bibnamefont {Sun}},
  \bibinfo {author} {\bibfnamefont {S.}~\bibnamefont {Friedemann}}, \bibinfo
  {author} {\bibfnamefont {S.}~\bibnamefont {Wirth}}, \bibinfo {author}
  {\bibfnamefont {C.}~\bibnamefont {Krellner}}, \bibinfo {author}
  {\bibfnamefont {S.}~\bibnamefont {Kirchner}}, \bibinfo {author}
  {\bibfnamefont {E.~M.}\ \bibnamefont {Nica}}, \bibinfo {author}
  {\bibfnamefont {R.}~\bibnamefont {Yu}}, \bibinfo {author} {\bibfnamefont
  {Q.}~\bibnamefont {Si}},\ and\ \bibinfo {author} {\bibfnamefont
  {F.}~\bibnamefont {Steglich}},\ }\bibfield  {title} {\bibinfo {title}
  {Interplay between unconventional superconductivity and heavy-fermion quantum
  criticality: {CeCu}\textsubscript{2}{Si}\textsubscript{2} versus
  {YbRh}\textsubscript{2}{Si}\textsubscript{2}},\ }\href
  {https://doi.org/10.1080/14786435.2018.1511070} {\bibfield  {journal}
  {\bibinfo  {journal} {Philosophical Magazine}\ }\textbf {\bibinfo {volume}
  {98}},\ \bibinfo {pages} {2930} (\bibinfo {year} {2018})}\BibitemShut
  {NoStop}%
\bibitem [{\citenamefont {Knappov\'a}\ \citenamefont{et~al.}()}]{Knappova2023}%
  \BibitemOpen
  \bibfield  {author} {\bibinfo {author} {\bibfnamefont {P.}~\bibnamefont
  {Knappov\'a}}\ \bibinfo {author} {\bibnamefont {et~al.}},\ }\bibfield
  {title} {\bibinfo {title} {Study of superconducting regimes in
  {YbRh$_2$Si$_2$} using dc-magnetisation and suceptibility},\ }\bibinfo {note}
  {in preparation}\BibitemShut {NoStop}%
\bibitem [{\citenamefont {Levitin}\ \citenamefont{et~al.}()}]{Levitin2022}%
  \BibitemOpen
  \bibfield  {author} {\bibinfo {author} {\bibfnamefont {L.~V.}\ \bibnamefont
  {Levitin}}\ \bibinfo {author} {\bibnamefont {et~al.}},\ }\bibfield
  {title} {\bibinfo {title} {Multiple superconducting transport regimes in
  heavy fermion metal {YbRh$_2$Si$_2$}},\ }\bibinfo {note} {in
  preparation}\BibitemShut {NoStop}%
\bibitem [{\citenamefont {Liu}\ \emph {et~al.}(2021)\citenamefont {Liu},
  \citenamefont {Chong}, \citenamefont {Sharma},\ and\ \citenamefont
  {Davis}}]{Liu2021}%
  \BibitemOpen
  \bibfield  {author} {\bibinfo {author} {\bibfnamefont {X.}~\bibnamefont
  {Liu}}, \bibinfo {author} {\bibfnamefont {Y.}~\bibnamefont {Chong}}, \bibinfo
  {author} {\bibfnamefont {R.}~\bibnamefont {Sharma}},\ and\ \bibinfo {author}
  {\bibfnamefont {S.}~\bibnamefont {Davis}},\ }\bibfield  {title} {\bibinfo
  {title} {Discovery of a {Cooper-pair} density wave state in a
  transition-metal dichalcogenide},\ }\href
  {https://doi.org/10.1126/science.abd4607} {\bibfield  {journal} {\bibinfo
  {journal} {Science}\ }\textbf {\bibinfo {volume} {372}},\ \bibinfo {pages}
  {1447} (\bibinfo {year} {2021})}\BibitemShut {NoStop}%
\bibitem [{\citenamefont {Fradkin}\ \emph {et~al.}(2015)\citenamefont
  {Fradkin}, \citenamefont {Kivelson},\ and\ \citenamefont
  {Tranquada}}]{Fradkin2015}%
  \BibitemOpen
  \bibfield  {author} {\bibinfo {author} {\bibfnamefont {E.}~\bibnamefont
  {Fradkin}}, \bibinfo {author} {\bibfnamefont {S.~A.}\ \bibnamefont
  {Kivelson}},\ and\ \bibinfo {author} {\bibfnamefont {J.~M.}\ \bibnamefont
  {Tranquada}},\ }\bibfield  {title} {\bibinfo {title} {Colloquium: {Theory} of
  intertwined orders in high temperature superconductors},\ }\href
  {https://doi.org/10.1103/revmodphys.87.457} {\bibfield  {journal} {\bibinfo
  {journal} {Reviews of Modern Physics}\ }\textbf {\bibinfo {volume} {87}},\
  \bibinfo {pages} {457} (\bibinfo {year} {2015})}\BibitemShut {NoStop}%
\bibitem [{\citenamefont {Weast}(1975)}]{Weast1975}%
  \BibitemOpen
  \bibfield  {author} {\bibinfo {author} {\bibfnamefont {R.}~\bibnamefont
  {Weast}},\ }\href@noop {} {\emph {\bibinfo {title} {Handbook of chemistry and
  physics : a ready{\textendash}reference book of chemical and physical
  data}}}\ (\bibinfo  {publisher} {CRC Press},\ \bibinfo {address}
  {Cleveland},\ \bibinfo {year} {1975})\BibitemShut {NoStop}%
\bibitem [{\citenamefont {Cowan}(2021)}]{Cowan2021}%
  \BibitemOpen
  \bibfield  {author} {\bibinfo {author} {\bibfnamefont {B.}~\bibnamefont
  {Cowan}},\ }\href
  {https://www.ebook.de/de/product/39956742/brian_cowan_topics_in_statistical_mechanics.html}
  {\emph {\bibinfo {title} {Topics in Statistical Mechanics}}}\ (\bibinfo
  {publisher} {WSPC (Europe)},\ \bibinfo {year} {2021})\BibitemShut {NoStop}%
\end{thebibliography}
%

\end{document}